\documentclass{WileyMSP-template}
\usepackage{amsmath,amstext,amssymb,mathtools,graphicx,color,xcolor,soul}
\usepackage[super]{cite}
\usepackage[colorlinks,urlcolor=blue,citecolor=gray]{hyperref}

\graphicspath{{./}{plots/}{template/}}

\definecolor{PineGreen}{HTML}{019286}

\newcommand{\citeref}[1]{{Ref.~[\citen{#1}]}}

\newcommand{\gcc}{{\textrm{g}\ \textrm{cm}^{-3}}}
\newcommand{\cmg}{\textrm{cm}^2\ \textrm{g}^{-1}}
\newcommand{\ergs}{\textrm{erg}\ \textrm{s}^{-1}}

\newcommand{\Msun}{{$M_\odot\ $}}
\newcommand{\msun}{{$M_\odot$}}

\makeatletter
\renewcommand\@citess[1]{\textsuperscript{[#1]}}
\makeatother

\begin{document}

\pagestyle{fancy}
\rhead{\includegraphics[width=2.5cm]{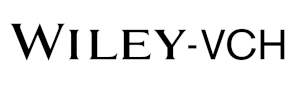}}

\twocolumn[{%
\title{
Heavy elements and electromagnetic transients from neutron star mergers}
\maketitle

\author{Stephan Rosswog$^{1,2}$ \; \& \;} 
\author{Oleg Korobkin$^3$}


\begin{affiliations}
$^1$Hamburger Sternwarte, University of Hamburg, Gojenbergsweg 112, 21029 Hamburg, Germany;\\ Email Address: stephan.rosswog@uni-hamburg.de\\
$^2$Department of Astronomy \& Oskar Klein Centre, Stockholm University, SE-106 91, Stockholm,\\ Sweden\\

$^3$Center for Theoretical Astrophysics, Los Alamos National Laboratory, Los Alamos, NM 87545,
USA\\
Email Address: korobkin@lanl.gov
\end{affiliations}

\keywords{r-process, neutron stars, gravitational waves, kilonovae}

\begin{abstract}
Compact binary mergers involving neutron stars can eject a fraction of their mass
to space. Being extremely neutron rich, this material undergoes rapid neutron 
capture nucleosynthesis, and the resulting radioactivity powers fast, short-lived 
electromagnetic transients known as kilonova or macronova. Such transients are 
exciting probes of the most extreme physical conditions and their observation 
signals the enrichment of the Universe with heavy elements.
Here we review our current understanding of the mass ejection mechanisms, 
the properties of the ejecta and the resulting radioactive transients. 
The first well-observed event in the aftermath of GW170817 
delivered a wealth of insights, but much of today's picture of such events is still
based on a patchwork of theoretical studies.
Apart from summarizing the current understanding, we also point out questions
where no consensus has been reached yet, and we sketch possible directions
for the future research. 
In an appendix, we describe a publicly available 
heating rate library based on the {\tt WinNet} 
nuclear reaction network, and we provide a simple fit formula
to alleviate the implementation in hydrodynamic simulations. \\
\end{abstract}

\medskip

}]

\justify

\section{Introduction}

The basic nuclear physics of the rapid neutron capture, or ``$r$-process'',
responsible for about half of the elements heavier than iron, has been 
laid out already in the 1950ies of the last century \cite{cameron57,burbidge57}. 
The astrophysical site where it actually happens, however, has remained 
a mystery for the decades to come. Throughout this time, the vast majority 
of the research community considered core-collapse supernovae as the 
most likely production site, despite the fact that such models were 
perpetually struggling to reach the 3rd $r$-process or ``platinum'' peak
around the nucleon number $A=195$.
For some time there was hope to produce such elements
in high-entropy winds driven from newborn proto-neutron stars 
\cite{woosley94,takahashi94,hoffman97,farouqi10}, but a slew of more recent 
supernova investigations does not support the emergence of the  very large entropies
that would be needed \cite{arcones07,roberts10,fischer10,huedepohl10,roberts12a,martinez_pinedo12a,martinez_pinedo14},
and today core-collapse supernovae are considered to contribute at best
to the light $r$-process elements. More extreme, and likely more rare, stellar 
explosions that magnetorotationally launch jets 
\cite{leblanc70,symbalisty85,cameron03,nishimura06,moesta15,nishimura17,moesta18,reichert21}, and also collapsars
\cite{pruet03,pruet04,surman06,siegel19a,siegel19b,miller20} have  been suggested as possible sites 
of heavy $r$-process, but while this is physically plausible, it is currently 
not clear, how robustly nature realizes the specific conditions  \cite{cameron03,surman04,siegel17a,miller20}
that are needed for heavy $r$-process in jet- and accretion-disk outflows  and 
how large the overall contribution from such sources is
on a cosmic scale.

The Milky Way is enriched by $r$-process elements at an average rate of
 roughly 1\Msun Myr$^{-1}$ \cite{qian00,rosswog17a,nakar20}, but this
 could, in principle, be the result of either frequent, low-yield (e.g. common
 core-collapse supernovae) or rare, high-yield events (e.g. neutron
 star mergers). A number of independent arguments ranging from large
 variance of $r$-process matter in metal-poor stars and ultra-faint
 dwarf galaxies \cite{beniamini16a,ji16,hansen17a,tsujimoto17} to
 deep-sea floor $^{244}$Pu \cite{wallner15a,hotokezaka15a} 
 all point to very rare events ejecting on average several percent of a
 solar mass and occurring at $\sim 1/1000$ of the core-collapse supernova
 rate, see e.g. \citeref{macias18,nakar20,cowan21,farouqi22} for more detailed discussions
 of this topic. In other words, also these arguments strongly disfavour
 regular supernovae as major source, consistent with modern supernova 
 simulations. Instead, they favor substantially rarer events such as 
 neutron star mergers or rare breeds of stellar explosions.

The idea that decompressed neutron star matter could be a promising $r$-process
site was discussed by Lattimer and Schramm in the context of neutron star
black hole mergers \cite{lattimer74,lattimer76} and further 
discussed in Symbalisty and Schramm \cite{symbalisty82}, this time considering
neutron star mergers. These
ideas were met with scepticism, both because no gravitational-wave dri\-ven merger
had, of course, been observed at the time, and because it is non-trivial to unbind matter from
a neutron star: the gravitational binding energy exceeds 150 MeV/\-baryon. The idea of decompressing neutron star matter
gained further credibility after the discussion paper of Eichler et al.\cite{eichler89}, which placed
neutron star mergers in a broader astrophysical perspective and connected them
not only with nucleosynthesis, but also with neutrino and gamma-ray bursts.
The first hydrodynamic plus nucleosynthesis calculations 
\cite{rosswog98b,rosswog99,freiburghaus99b} showed that neutron star mergers
eject $\sim$ 1\% of a solar mass per event, that these extremely neutron-rich
ejecta effortlessly reach the platinum peak and beyond, and that, when
folded with the estimated merger rates, the produced $r$-process mass is  enough
for neutron star mergers being
the major (or maybe even the only) $r$-process source.

While already these studies made a strong case for neutron star mergers as important
$r$-process site, the picture has become substantially more detailed in recent years. 
On the one hand more ejecta channels were identified: dynamic ejecta launched by shocks (rather
than tidal torques) \cite{oechslin07,bauswein13b,hotokezaka13,radice18a,nedora21}, neutrino-
and magnetically driven winds \cite{dessart09,perego14b,siegel14a,fahlman18,fujibayashi20}, (magnetohydrodynamic and nuclear) unbinding of accretion tori \cite{metzger08a,beloborodov08,lee09,siegel17a,siegel18,fernandez19,miller19}, and
last but not least, by viscous processes\cite{radice18b,fujibayashi18,fujibayashi20} in the rem\-nant. 
 On the other hand, it was realized that these different ejection channels 
go along with different weak interaction histories and therefore the ejecta have different
neutron-to-proton ratios. This is usually quantified via the ``electron fraction''  $Y_e= n_e/(n_n+n_p)= n_p/(n_n+n_p)$, where the indexed $n$'s refer to the number densities of electrons, 
neutrons and protons. Under the conditions realized in neutron star mergers, $Y_e$ has
{\em the} decisive impact: the resulting $r$-process abundance pattern changes rather
abruptly\cite{korobkin12a,lippuner15,rosswog18a} around $Y_e\approx 0.25$, with lower values 
producing ``heavy $r$-pro\-cess'' with nucleon numbers $A > 130$ while values above this 
threshold produce only light ``$r$-process'' ($A<130$).\\
Tidal dynamical ejecta are hardly heated and expand rapidly, so that
weak interactions have no chance to substantially change the ejecta's number of neutrons 
and protons and they keep their original, cold $\beta$-equi\-librium 
electron fraction $Y_e\sim 0.05$ and therefore produce strong $r$-process along 
the lines found in \citeref{freiburghaus99b}. 
Shock-heated dynamic ejecta, in contrast, are substantially
heated and, although they escape with large velocity, their electron fraction may be
substantially altered. Neutrino-driven winds start out with low velocities in an
environment flooded with neutrinos, and as they become secularly driven out of
the remnant's gravitational pull, they have plenty of time to change their 
neutron-to-proton ratio.  The detailed electron fractions in most of the more recently identified ejection channels are currently still unsettled questions 
\cite{siegel14a,siegel17a,siegel18,nedora19,fernandez19,miller19}, but it seems
clear that a much broader range than initially thought is realized in nature
and several studies found neutron star mergers to produce the whole range of $r$-process elements
\cite{wanajo14,just15,fernandez16b} rather than only the main $r$-process elements
with $A>130$ as found in the first study of \citeref{freiburghaus99b}.\\
 A number of excellent reviews related to the broad topic of our paper have appeared
in recent years. For example, with a strong focus on $r$-process, there are the
reviews of \citeref{sneden08,cowan21,siegel22}, \citeref{metzger19a,nakar20,margutti21,pian21} discusses
in detail the electromagnetic counterparts of compact binary mergers, while 
\citeref{shibata19a} and \citeref{radice20} focus more on the merger itself.

\begin{figure*}
  \centerline{\includegraphics[width=0.8\linewidth]{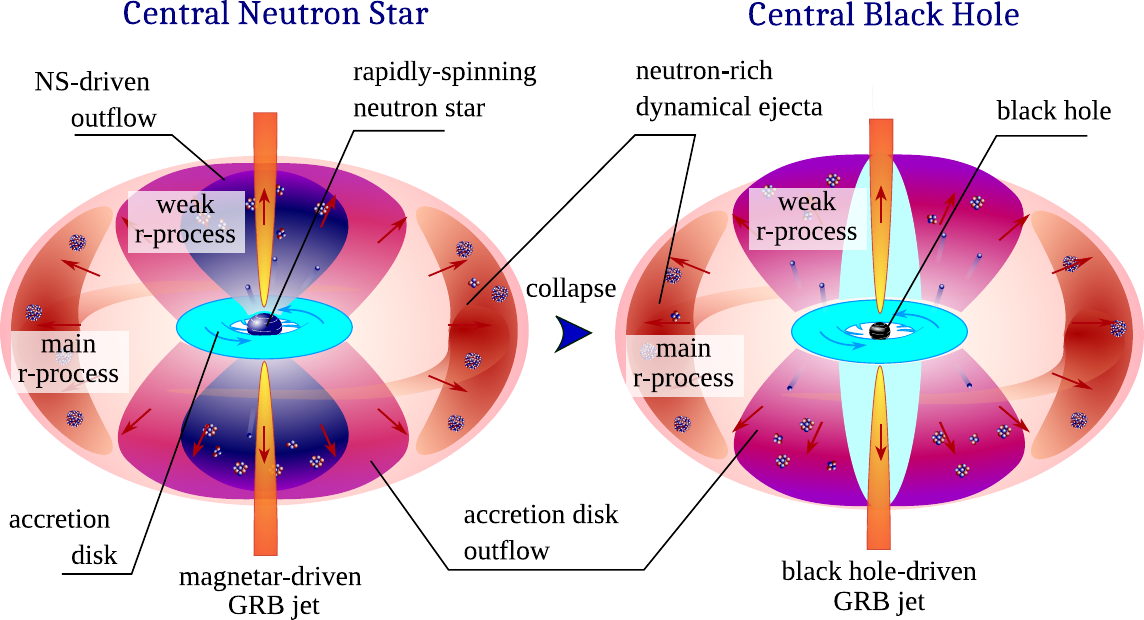}}
  \medskip \caption{Sketch of potential post-merger remnant structures.}
  \label{fig:remnant-sketch}
\end{figure*}

\section{The theoretical picture}
The first detection of a gravitational wave signal from a 
neutron star merger\cite{abbott17b} (GW170817) and its subsequent electromagnetic
transient \cite{abbott17c} (AT2017gfo) has overall provided a strong confirmation 
of the theoretical picture that had been developed by the theoretical
astrophysics community over the preceding two decades. Here we will
provide a short overview over the, observationally informed 
but predominantly theoretical, picture that has emerged over
the years.
It is worth mentioning that none of the studies that this picture is based upon 
contains all the relevant physics ingredients to sufficient accuracy,
and no study can claim full numerical convergence of the post-merger stage. What
we will present is also idealized in the sense that 
what we list below as separate channels  may in reality
occur in a concerted way. 
For example, winds may be driven by the combined action of 
neutrino-absorption and the dissipation of magnetic fields 
in a ``hot corona'' above the disk. In line
with the topic of this article, we will focus on the 
ejecta and the electromagnetic emission they cause.
But before we get there, we have to briefly summarize 
the dynamical evolution of a neutron star merger and outline
different pathways for its further evolution.

\subsection{Dynamical evolution}
The post-merger multi-messenger signatures --gravitational waves, neutrino
emission, mass ejection and electromagnetic emission-- are closely tied
to the fate of the merger remnant. Binary systems with masses 
$>2.8$ \Msun can collapse on a dynamical time scale (``no bounce'') to black holes, with
the exact threshold mass depending on the equation of state (EOS) \cite{hotokezaka11,bauswein13b,koeppel19,bernuzzi20b,kashyap22}.
Black holes formed in such a prompt collapse are not surrounded by 
massive disks unless the mass ratio of the  binary, $q$, deviates
substantially from unity.  It takes 
$q\lesssim 0.8$ to form disks of moderate mass ($>0.01$ \msun)
\cite{shibata06c,kiuchi09,hotokezaka13b}, for even more extreme mass ratios
the tidal shredding of the secondary neutron star can still lead to sizeable
disk masses \cite{bernuzzi20b}. For the close-to-equal mass
case there are hardly any ejecta expected and such systems may be 
electromagnetically undetectable. For mass ratios smaller
than about 0.8 the ejecta mass can exceed 10$^{-3}$ \Msun in a prompt
collapse.

Out of the 13 galactic neutron star binaries with total masses published in
\citeref{tauris17}, 11 have a mass below 2.8 \msun. It is therefore
likely that the majority of neutron star binaries goes at least
temporarily through a phase where a massive neutron star is present. 
This has a major impact since the central remnant can keep emitting
gravitational waves, shed mass into a torus, eject matter, and
provide a strong central neutrino source that can modify the 
neutron-to-proton ratio. 
For this case with a central neutron star remnant, the amount of ejecta
is strongly dependent on the EOS. If the EOS is stiff, neutron stars of
a given mass have larger radii (than for a soft EOS) and therefore the
tidal effects are larger, and the merger occurs earlier, at larger orbital separations, and with smaller
orbital velocities. In addition, the sound speed of  stiff EOSs is larger, so
it is more difficult to shock the neutron star matter. Therefore, shock heating
is not very efficient, and the post-merger oscillations are less
violent, so not much mass is ejected dynamically. Softer equations of
state, in contrast, lead to a longer inspiral, larger velocities at
contact, stronger shocks, more violent post-merger oscillations and
therefore larger amounts of dynamically ejected mass. However, even for soft EOSs
the dynamic ejecta mass from equal-mass binaries is typically only 
a few times 
$\sim 10^{-3}$~\msun\cite{hotokezaka13a,bauswein13a,radice18a,rosswog22b},
while unequal-mass binaries produce larger amounts 
of ejecta with a more pronounced tidal contribution 
and lower electron fractions. Extreme cases with very large mass ratios
and (implausibly) stiff EOSs\footnote{The used MS1b EOS is disfavoured
by the limits on tidal deformability \cite{abbott17b}.} can reach 
ejecta masses of $\sim 0.06$ \Msun 
of tidal ejecta \cite{dietrich17}.

\subsection{Ejection channels}
We sketch in Figure \ref{fig:remnant-sketch}  possible post-merger configurations 
together with various ejection channels. The left half of the figure illustrates the case
where a central neutron star survives, the right is a sketch of the configuration 
after the central neutron star has collapsed into a black hole or, alternatively, it could  be
produced in the merger of a neutron star with a stellar-mass black hole. 
 Keep in mind that neutron star mergers with nearly equal masses
may not form  any torus of significant mass. If a neutron star survives either 
temporarily or indefinitely, it will be very highly rotating and likely have a 
beyond-magnetar field strength \cite{price06,kiuchi15,palenzuela22}. This may 
continuously inject a large flux of energy into the remnant which could potentially
lead to electromagnetic light curves that are very different from the case 
where the central object is a black hole.\cite{wollaeger19}\\
As alluded to above, the number of known ejecta channels has substantially 
increased in recent years, and we group them here broadly into ``dynamic ejecta'',
``winds'', and ``torus ejecta'', which we discuss separately below.

\subsubsection{Dynamic ejecta}
The historically first identified ejection channel are
so-called dynamic ejecta that are launched on a dynamical
time scale ($\sim 1$ ms) at merger. The ejecta of equal
mass systems are dominated by {\em shock-driven} 
ejecta, while with decreasing mass ratio $q$ the contribution 
from {\em tidal ejecta} becomes more and more important. 
The shock-driven ejecta result from
both matter being shocked at the interface between the merging
stars, and from strong post-merger pulsations of
the remnant when outward travelling sound waves steepen
into shocks as they pass through matter of decreasing
density (and therefore decreasing sound speeds). Since
this matter becomes heated to temperatures of a few MeV,
where the weak interaction time scales compete with the dynamical ones, 
its electron fraction can increase substantially by both positron captures, 
$n + e^+ \rightarrow p + \bar{\nu}_e$, and neutrino absorptions,
$n + \nu_e \rightarrow p + e^-$. The latter  process, however, needs
some time to ``ramp up'' since the neutrino luminosities
only reach large values once a substantial torus has been
assembled around the central object which typically takes $\sim 5$ ms.  
Tidal ejecta, in contrast, usually do not experience shocks and are 
therefore ejected with their pristine, neutron star electron fraction of $Y_e\sim 0.05$ (but see the discussion in Sec.~\ref{sec:ejecta_interactions}). The detailed ejecta properties depend on the 
total mass, the mass ratio $q$ and the equation of state, 
but one can expect a broad distribution of $Y_e$ 
between $\sim 0.05$ and $0.4$ \cite{wanajo14,sekiguchi16a,radice18a}. The total amount 
of dynamic ejecta  found in simulations ranges from $\sim 10^{-4}$ to $\sim 10^{-2}$ \Msun \cite{rosswog99,hotokezaka13a,bauswein13a,rosswog13b,sekiguchi16a,lehner16a,dietrich17,radice18a,rosswog22b}
with soft equations of state typically ejecting more matter than stiff ones.

The average ejecta velocities are in a range between $\sim 0.15c$ to $\sim 0.3c$
\cite{hotokezaka13a,sekiguchi16a,radice18a,rosswog22b}, 
but for specific cases\footnote{These latter simulations use the conformal 
flatness approximation to General Relativity.} 
they can reach $\sim 0.4c$ \cite{bauswein13a}. 
Many groups have found small amounts of ejected matter  (the exact numbers are rather uncertain, but roughly of order $\sim 10^{-5}$ \msun) that 
reach asymptotic velocities in excess of 0.5$c$ \cite{hotokezaka13a,kyutoku14,bauswein13a,metzger15a,radice18a,dean21,rosswog22b,combi22}. While none 
of the groups can claim full numerical convergence of the properties of this small amount 
of mass from 3D simulations, it seems increasingly clear that such a high-velocity tail does exist. As Fig.~\ref{fig:ejectavelocity}
suggests, this high-velocity component is (at least
partially) due to shock-heated matter from the interface between the two neutron stars.
If indeed such a high-velocity component is
present, the ejecta should contain a significant fraction of free neutrons at freeze-out, 
and their decay should cause an early blue/UV transient on a time scale of several minutes 
to an hour \cite{kulkarni05,metzger15a}. Moreover, such a high-velocity
component would have a strong effect on the shock breakout signal\cite{beloborodov20} and it
could additionally cause synchrotron emission many months after a neutron star merger \cite{nakar11a,mooley17,hotokezaka18,hajela22}.\\
The majority of simulation studies has focussed on irrotational binary systems. This is because the combination of low viscosity in the
neutron star matter and the very short time ($\sim 0.01$s) the binary 
components interact tidally do not allow for a substantial tidal spinup \cite{bildsten92,kochanek92}. Nevertheless, there might be some residual
stellar spins left at merger and a few studies have explored their impact. In extreme cases,
spins can enhance the dynamic ejecta (and disk) masses by more than an order of magnitude \cite{papenfort21} and they can impart asymmetries to the
ejecta and kicks to the central remnants \cite{rosswog00,chaurasia20}.
\begin{figure*}[htp!]
  \centerline{\includegraphics[width=\textwidth]{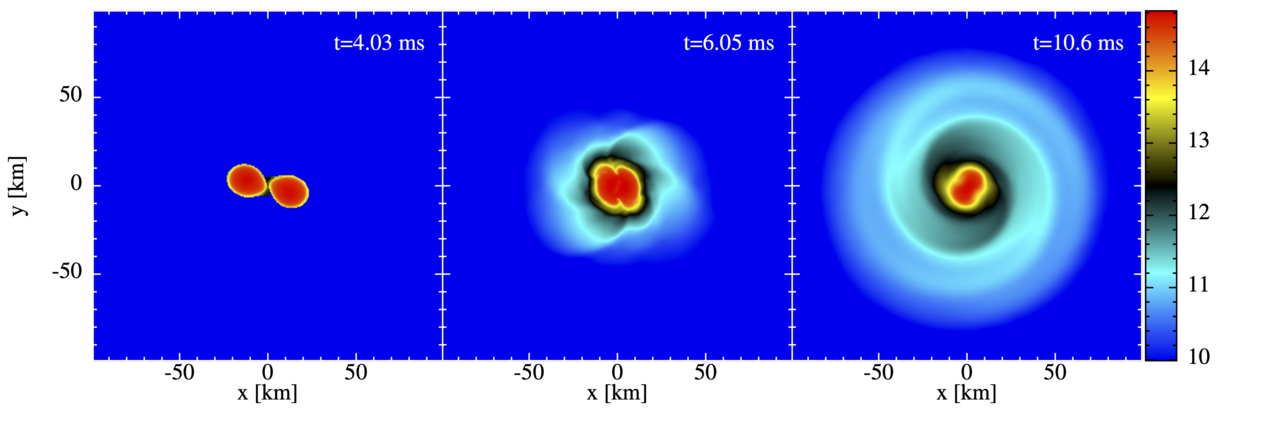}}
  \medskip
  \caption{Simulated merger of a  $2 \times 1.3 $ \Msun neutron star 
  binary with the MPA1 equation of state. The simulation was performed 
  with the Lagrangian Numerical Relativity code 
  {\tt SPHINCS\_BSSN} \cite{rosswog21a,diener22a,rosswog22a,rosswog22b}.}
  \label{fig:MPA_sequence}
\end{figure*}
\begin{figure*}[htp!]
  \centering
  \includegraphics[width=\textwidth]{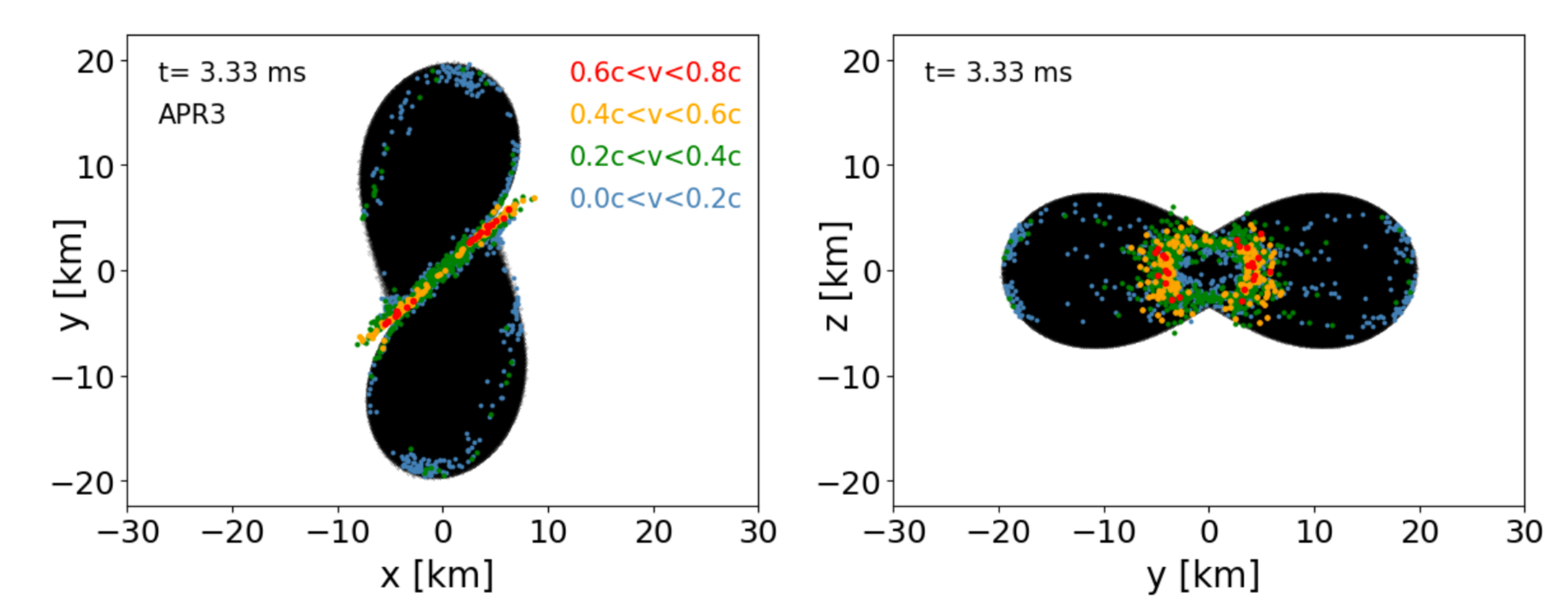}
  \medskip
  \caption{Particle positions (black) during the time of merger 
  (2$\times$ 1.3 \msun, APR3 equation of state). Overlaid are
  the positions of particles that are finally ejected with asymptotic 
  velocities $v_\infty$ between 0 and 0.2$c$ (blue), between 0.2 and 
  0.4 $c$ (green), between 0.4 and 0.6$c$ (orange) and above 0.6$c$ (red). 
  The simulation was performed with the Lagrangian Numerical 
  Relativity code {\tt SPHINCS\_BSSN} \cite{rosswog21a,diener22a,rosswog22a,rosswog22b}.}
  \label{fig:ejectavelocity}
\end{figure*}

\subsubsection{Winds}
On a time scale exceeding the dynamical one, various processes can unbind matter from
the accretion disk\footnote{Sometimes the word ``disk''
is reserved for geometrically thin accretion disks and ``torus'' for thicker matter configurations
that are thermally ``puffed up'' because they cannot efficiently cool on a dynamical time scale. 
A good fraction of the post-merger accretion may proceed in a neutrino-cooled regime, so one may argue
that "disk" might be more appropriate. However, we do not want to distinguish in our wording between the
different cases and simply use the word "torus".}. First,
and especially for the likely common case that a central neutron star survives at least
temporarily, energy deposition by neutrino absorption can inflate the disk vertically and
can drive quasi-spherical outflows on time scales of $\sim 100$ ms \cite{dessart09,perego14b,radice18a}. 
Since this matter is exposed to intense neutrino irradiation, and the time scales are 
relatively long, its neutron-to-proton ratio can change substantially and will, starting
from the pristine $Y_e$-value $<0.1$, evolve towards 
an equilibrium value \cite{qian96b,martinez_pinedo12} of
\begin{equation}
Y_e^{\rm eq} \approx \left(1 + 
\frac{L_{\bar{\nu}_e}}{L_{\nu_e}}
\frac{W_{\bar{\nu}_e}}{W_{\nu_e}} 
\frac
{\epsilon_{\bar{\nu}_e} - 2 \Delta + 1.2 \Delta^2/\epsilon_{\bar{\nu}_e}}
{\epsilon_{\nu_e} + 2 \Delta + 1.2 \Delta^2/\epsilon_{\nu_e}}
\right)^{-1},
\end{equation}
where the $L_i$ denote the neutrino luminosities, $W_i= 1+ \eta_i \langle E_i \rangle/m_bc^2$ are weak magnetism correction factors with $\eta_{\nu_e}=1.01$
and $\eta_{\bar{\nu}_e}=-7.22$, $\Delta= (m_n-m_p)c^2\approx$ 1.29 MeV 
and $\epsilon_i$ are the ratios between the average squared neutrino energy
and the average neutrino energies (typically $\epsilon_\nu \approx 1.2 E_\nu$).
For the typical neutrino field properties reached in a merger, this equilibrium
value is $\sim0.4$, see e.g. Fig. 23 in \citeref{farouqi22}.
It is, however, by no means guaranteed that during the dynamic
evolution of the remnant the ejecta will reach this equilibrium value. Numerical studies show
that a broad distribution of electron fractions between $\sim0.1$ and
$\sim0.4$ and, for long-lived central remnants, ejecta masses of $\sim 0.01$ \Msun are reached \cite{dessart09,perego14b,martin15,radice18a}.

Apart from this neutrino-driven wind, the accretion process itself will
unbind some material: as neighbouring, quasi-circular disk segments couple
viscously, angular momentum is removed from the inner regions, thus allowing 
accretion, and it is transferred to the outer portions of the flow which, by 
being continuously fed with angular momentum, can expand quasi-spherically to infinity. Another, 
likely more efficient mechanism is realized by the so-called 
``spiral wave winds'' \cite{nedora19,nedora21}. 
They emerge in cases where the central remnant
does not collapse quickly to a black hole and 
instead forms non-axisymmetric bar modes that continuously pump angular momentum into
the surrounding accretion disk. As a result,
matter becomes unbound,  predominantly in the orbital plane, in the form
of a spiral wave. This process continues until either the central remnant
collapses or the bar mode is dissipated (by gravitational waves or other mechanisms). 

If large-scale, ordered magnetic fields emerge in the remnant, they also can drive outflows.
Both the spiral-wave and the magnetically driven winds are expected to reach
masses of $\sim 0.01$ \msun, asymptotic velocities of 0.1-0.2c and have electron fractions
in excess of the critical value of $\approx 0.25$ \cite{korobkin12a,lippuner15}. 
With such properties, they are viable candidates \cite{nedora19,ciolfi20} 
to explain the early, blue kilonova component of the transient that 
followed GW170817, see e.g. \citeref{evans17,nicholl17,mccully17}. Depending on magnetic field 
strength and rotational period, magnetized winds can reach larger 
velocities \cite{metzger18a} than spiral wave winds which might 
potentially help to explain the large inferred velocities of the blue component  of GW170817.

\subsubsection{Torus ejecta}
While the above discussed ejecta components could potentially provide
masses close to what was estimated for GW170817 (but keep in mind that the 
current estimates could be too large due to the assumption of sphericity \cite{korobkin21}), 
it is worth noting that the winds (neutrino-, magnetic- and spiral wave-driven) are all
candidates for the blue kilonova component.
At the same time, the red component was estimated in some studies
to be the result of as much as $\sim 0.04$ \Msun of matter with a rather low average
velocity of $\sim 0.1c$ (e.g. \citeref{cowperthwaite17}). Dynamic ejecta are
the only channel discussed so far that could  produce a red signal, but they 
fall short in producing this large amount of mass and their average velocity 
is substantially in excess of the inferred value of 0.1$c$. A channel that 
potentially can provide ejecta with such  properties is the late-time ($\sim 1$~s) 
unbinding of the accretion disk 
\cite{metzger08a,beloborodov08,siegel17a,siegel18,miller19,fernandez19} 
(but see the discussion below for possible caveats).

The torus ejection mechanism is a combination of two effects: as the disk expands viscously, 
it cools, and, once the temperature drops below $\sim 5\times 10^9$ K, the neutrons and
protons that constitute the disk material combine into $\alpha$-particles, releasing
7.1 MeV/nucleon. From the location where this happens, this energy release alone
would launch  ejecta with
asymptotic velocities of $\sim 0.05c$. Detailed numerical simulations of the accretion 
of thick disks (initially in equilibrium) that account for the assembly of nucleons 
into $\alpha$-particles and the evolution of magnetic fields \cite{siegel17a,siegel18}, 
find that steady-state MHD turbulence develops strong magnetic fields in the disk and 
generates a hot disk corona that, helped by the assembly of nucleons into $\alpha$-particles, 
launches powerful thermal, quasi-spherical outflows. Via this mechanism up to $\sim 40\%$
of the initial disk mass can become unbound with the outflows reaching asymptotic 
bulk velocities\footnote{The outflows also contain high-velocity tails  up to $\sim 0.5c$; J. Miller, private communication.} of $\sim 0.1c$.  Thus, a typical accretion disk from a neutron star merger 
with $\sim 0.1$ \Msun as initial disk mass, would
plausibly produce the masses and outflow 
velocities that have been inferred for the red component.

Interestingly, at the expected accretion rates (which are those 
that are needed to launch a GRB),
the inner parts of the disk mid-plane regulate themselves
into a state of mild electron degeneracy which creates a
negative feedback cycle resulting in electron fraction
values of ${Y_e\approx0.1}$ in the inner disk regions
\cite{beloborodov03,kawanaka07,chen07,beloborodov08,metzger09b,siegel18}. 
This is interesting for at least two reasons.
First, it deviates from the maybe na\"ive expectations that 
a hot, but initially very neutron-rich disk should increase
all of its electron fraction rapidly (in some parts of the disk
this happens, though), and it provides a rather robust mechanism
to produce low-$Y_e$ material. If the electron fraction of 
this material should rise only mildly during the ejection process,
disk ejecta below the critical value $Y_e^{\rm crit}\approx 0.25$
\cite{korobkin12a,lippuner15} for heavy $r$-process 
production can emerge. Second, there is an interesting 
potential connection to observed ``actinide boost stars''.
These are stars that have thorium- or uranium-to-europium 
ratios that substantially exceed the solar values 
\cite{roederer10,holmbeck18,holmbeck19a}. Nuclear parameter
studies \cite{wu17,eichler19,holmbeck19b} that explored under
which conditions such ``actinide boosts'' can occur found that 
they require a rather narrow electron fraction range of
$0.1 < Y_e < 0.15$. The only astrophysical ``engines'' that we are 
aware of that have good physical reasons to preferentially 
produce electron fractions in this range are such accretion 
disks. It has therefore been speculated
\cite{farouqi22} that such disks might be the sites that produce
the abundances for the ``actinide boosts'' in stars. If true, this
would also connect actinide boost stars to the sources of gamma-ray bursts.

The picture of accretion disks ejecting low electron fraction material, 
being a good candidate for the red kilonova component of GW170817 
and a major production site of 3rd-peak $r$-process, however, 
is not undisputed. Recent studies by Miller and collaborators \cite{miller19,miller19b,miller20} 
evolving accretion disks in  fixed Kerr spacetimes employing Monte Carlo 
neutrino transport comes to a different conclusion, namely that the ejecta should
be predominantly above $Y_e= 0.25$ and therefore produce a blue kilonova.
Given the (physical and numerical) complexity of the challenge, it is
not entirely surprising that at this early stage of studying such 3D-GRMHD 
accretion disks with neutrino cooling the results do not agree. 
However, it is worth keeping in mind that  the studies of Siegel and collaborators and Miller and collaborators investigate different 
physical parameters and use different approximations of the involved physics, 
in particular in their treatment of neutrinos. The latter has been studied in 
detail by \citeref{just22}, though not in GR, but using Newtonian $\alpha$-disk/
MHD simulations with a pseudo-potential for the black hole\cite{artemova96}. 
The authors point out the importance of the initial disk electron fraction 
(of which some parts can become unbound without much neutrino interactions),
the substantial differences between an $\alpha$-disk and an MHD treatment, 
but they also stress that plausible  simplifications such as 
neglecting electron masses and the mass difference between neutrons and
protons can have an impact. It seems in particular that neutrino absorption effects
become important already for disks of moderate masses ($>0.01$~\msun)
\cite{fernandez20,just22} and that relatively simple treatments
such as leakage schemes or an M1-approach may underestimate
the related effects.
To the best of our knowledge, the issue of the 
composition of torus ejecta is at the time of 
writing not yet settled and possible solutions include the possibility
that all the differences are due to the neutrino treatment, but 
it could also be that different
physical parameters (black hole masses and spins, disk masses, isolated
black hole disk systems as they may emerge after a neutron star merger
vs continuously fed systems as they may emerge in collapsar case)
can lead to different physical outcomes. Clearly,
substantially more work is needed to settle this issue in the future.

In passing, we want to stress that {\em if} the resolution of this
issue should be the neutrino transport and the torus ejecta should be
above $Y_e= 0.25$ and {\em if} the red ejecta mass is really $\sim 0.04$ \Msun 
(but see our discussion below), then we would lack a convincing explanation of the red 
kilonova, since tidal ejecta are challenged to produce the needed large amount
of mass.\footnote{Some extreme models with both an extreme mass ratio of $q\approx 0.5$
and the extremely stiff MS1b EOS could produce such large ejecta masses \cite{dietrich17}.
The MS1b EOS, however, is strongly disfavoured by the 
LVC results on the tidal deformability in GW170817 \cite{abbott17h}.}

\begin{figure}
    \centering
    \includegraphics[width=\linewidth]{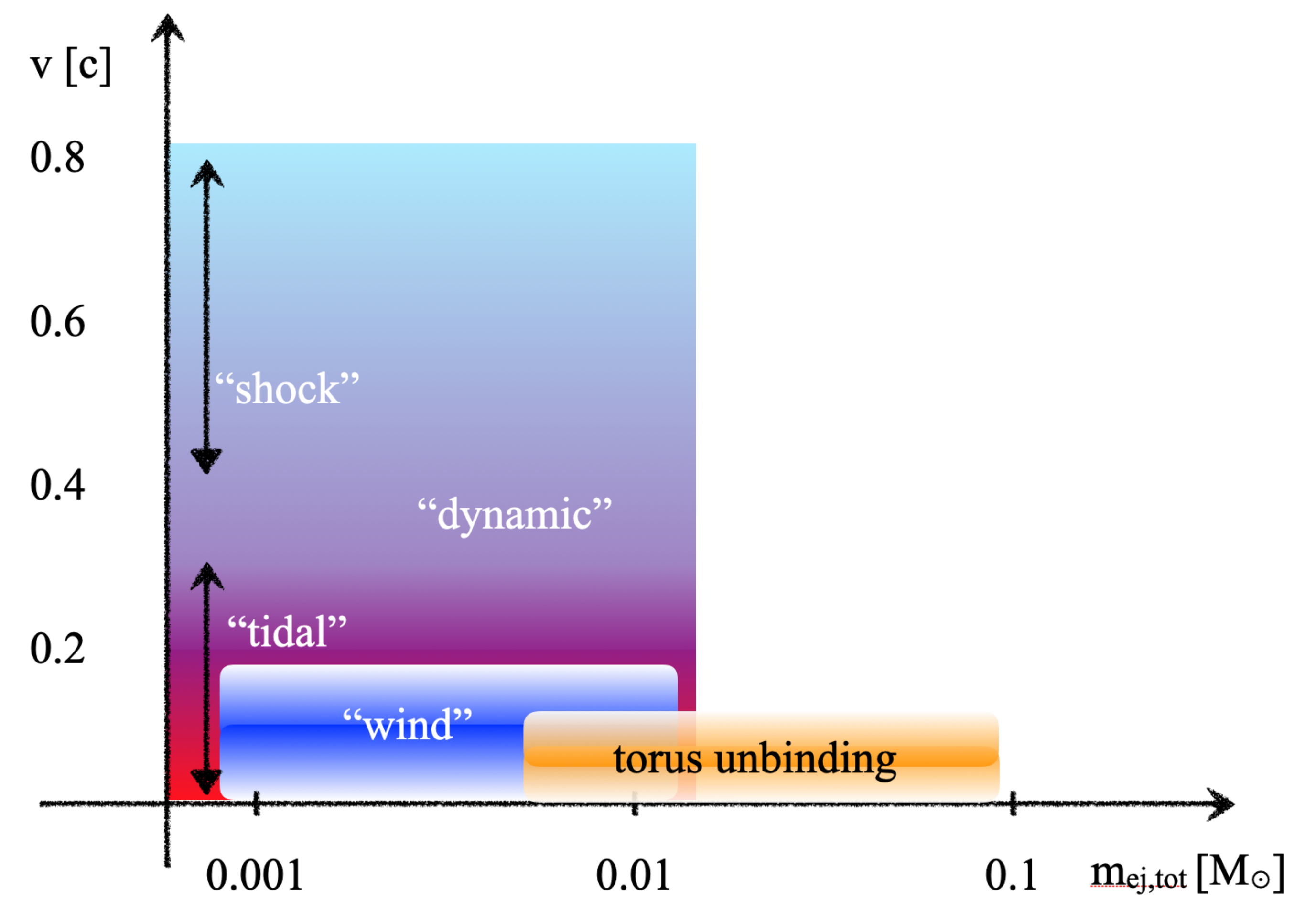}
    \caption{Sketch of the total masses and expected velocity ranges
    in different ejecta channels.}
    \label{fig:m_v_ejecta}
\end{figure}

We sketch in Fig.~\ref{fig:m_v_ejecta} the currently expected ranges 
of total ejecta mass and velocity.

\subsection{Radioactively powered transients}

Partially sparked by a conference proceedings article
showing the first nucleosynthesis calculations for neutron 
star merger ejecta \cite{rosswog98b}, Li \& Paczy\'nski \cite{li98}
explored in a simple model what a neutron star merger's 
electromagnetic emission due to radioactive ejecta  may 
look like. They realized that due to the small
ejecta mass and the rapid expansion velocity a short-lived
supernova-like transient should result. As in a thermonuclear supernova,
the bulk of the initial internal energy is consumed in the 
optically thick, initial expansion phase and detectable 
emission can only be expected, if at the time when 
$t= \tau_{\rm diff}$, the latter being the diffusion time, 
a (radioactive) heating source is present. 
Adopting an effective gray opacity of $\kappa=0.2\ \cmg$, 
Li \& Paczy\'nski predicted a one-day lasting blue transient 
with a peak luminosity of over $10^{44}\ \ergs$. 
Metzger et al. in \citeref{metzger10b}
improved on this simple model by using nuclear network 
calculations to determine the nuclear heating rate and by applying 
1D radiative transfer to calculate light curves. Using (the too low) gray opacity 
representative for iron group elements, they found 
transients peaking after $\sim 1$ day in the optical, 
but with luminosities that are more than 
two orders of magnitude lower than found by
Li \& Paczy\'nski \cite{li98}.

Soon after, it was realized \cite{kasen13a,tanaka13a} 
that ejecta consisting of $r$-process elements, in particular 
if they include lanthanides and actinides, contain 
orders of magnitude more optical transition lines than 
the iron-group elements. In an expanding medium, such 
as merger ejecta, these lines dominate the material 
opacity, raising it by a factor of $\sim 10$ for the 
light $r$-process, and a factor of 100 to 1000 for 
lanthanides and actinides \cite{kasen13a,tanaka13a,fontes15}. 
Thus, if heavy $r$-process is present, one expects a more 
modest peak luminosity of $\sim10^{41}\ \ergs$ and a transient
duration of about a week. Soon after these new insights on the opacity, 
it became clear that the various ejecta of neutron star mergers contain 
a broad range of electron fractions \cite{wanajo14,just15,fernandez16b}
so that both a ``red'' and a ``blue kilonova'' are plausible 
possibilities.

The calculations of \citeref{kasen13a,tanaka13a} predicted
broad spectral features, but no individual lines due to 
large expansion velocities. Of all the lanthanides and 
actinides, neodymium is expected to dominate the
opacity~\cite{kasen17,fontes17a, even20}. Early spectra 
appear as an almost featureless blue continuum blackbody 
that rapidly shifts to longer wavelengths, peaking in 
near-infrared on a timescale of a week~\cite{wollaeger18}.
Later spectra strongly deviate from blackbody, showing 
broad spectral features and a sharp cutoff at shorter
wavelength~\cite{wollaeger18,fontes20}. 

Effects from matter not being in local thermodynamic equilibrium
(``NLTE effects'') are expected for both early  and 
 late spectrum individual lines, so
a correct prediction of spectra requires NLTE effects 
to be accounted for~\cite{hotokezaka21,pognan22a}.
A detailed analysis of the cooling mechanisms shows 
that the temperature is supposed to remain approximately 
constant for an extended period of time, providing an
unchanged spectrum for several weeks~\cite{pognan22a, pognan22b}. 
NLTE effects in kilonova are only being tackled now 
and they are currently an area of active research.

\section{Confronting  observations}

While the above outlined picture has, despite its short\-comings, 
a sound physical basis, only some of the described aspects are
observationally confirmed.
What {\em are} the observational facts that we can consider as firmly established?
We will review them here as a guidance to understand where our
theoretical picture needs to be refined further in future work.
What we discuss here is certainly biased by our incomplete knowledge 
of the rapidly growing literature on the topic.

The visible and IR emission that followed the first observed neutron 
star merger GW170817 showed an early blue emission in the first 2-3 days, 
followed by a longer-lived red transient that decayed on a time scale 
of about a week \cite{evans17,nicholl17,mccully17,chornock17,pian17,tanvir17}. While rapidly rising initial blue 
emission is inconsistent with high-opacity ``main'' $r$-process (including 
lanthanides and heavier nuclei), blue emission is exactly what is expected
from radioactively powered light $r$-process ejecta \cite{martin15}. The 
relatively large ``blue'' ejecta masses and velocities (${\sim 0.025}$ \Msun 
and ${\sim 0.3c}$, see \citeref{kasen17}), however, are larger than 
what is expected from neutrino-driven winds alone. 
They could be accommodated, e.g. by the above discussed spiral-wave winds, 
provided that the central remnant survives for a few hundred milliseconds. 
One possible explanation for the high expansion velocity of the blue 
component is that we are actually observing fast-moving dynamical ejecta, 
which is reprocessing the flux from deeper, slower-moving layers~\cite{kawaguchi18}.
Light $r$-process elements, however, are not the only possible 
explanation, other suggestions for the origin of the blue 
component include cocoon shock cooling \cite{piro18}, 
or magnetar-like central engine activity
\cite{kisaka16,metzger18a}.

A longer lasting, redder component, on the other hand, is 
consistent with high-opacity, lanthanides-rich ejecta
\cite{tanvir17,troja17,kasen17}. The overall observed 
energetics, i.e. the bolometric luminosity, of AT2017gfo 
fits very well with the nuclear heating due to decaying 
$r$-process elements~\cite{kasliwal17,rosswog18a}. While 
we consider this as strong evidence for the production 
of $r$-process elements, it is not particularly constraining
in terms of which elements are involved, since a broad 
range of electron fractions, ${Y_e\sim0.05-0.35}$, produces nearly 
the same, power-law like heating (see Appendix~\ref{app:heating}).

The exact composition is more difficult to infer.
Spectral calculations were used to interpret features in the spectrum of
AT2017gfo as being due to the presence of strontium,
a light $r$-process element from  the first $r$-process peak
\cite{watson19,domoto21,gillanders21}. The presence of Sr would imply 
that the blue component of the ejecta were rather proton-rich, with
$Y_e>0.4$~\cite{domoto21}, while at the same time the bolometric 
luminosity suggests that the bulk of the ejecta should have 
$Y_e\le 0.35$~\cite{rosswog18a}. Given that the
pre-merger neutron stars contain only a very small mass fraction with
$Y_e>0.1$ ($\sim 10^{-4}$; see Fig.~23 in \citeref{farouqi22}), this
also underlines the pivotal role that weak interactions play in a
neutron star merger.

While the presence of Sr is plausible, its unique identification 
is a very challenging task and the so far collected evidence cannot be
considered as conclusive\footnote{We are grateful to Eli Waxman for 
sharing his detailed viewpoints on this topic with us.}. The 
challenges arise on the one hand from a serious lack of relevant and reliable
atomic data, in particular for elements in the second $r$-process peak \cite{gillanders22}.
On the other hand, existing studies have a large number
of free parameters out of which some are fixed ``by hand''
to plausible values while others are varied to obtain 
agreement between simple models and observations. This raises,
of course, questions about the uniqueness, robustness, and 
consistency of the results\footnote{For example, 
\citeref{gillanders22} find best-fit density values that 
are substantially below those found in the hydrodynamic simulations
they use to determine abundances. Such low values may actually be 
challenged to reproduce the minimum mass that is needed to explain 
the bolometric luminosity of AT2017gfo, $m_{\rm ej} > 0.015$ \msun \cite{rosswog18a}.}.
Helium has been considered as an alternative explanation for a spectral 
feature at 8000~\AA, but no conclusive evidence was found~\cite{perego22}.
A search for some of the third $r$-process peak elements such as 
Pt and Au in the photosphere was performed~\cite{gillanders21}, 
but with negative results.

The week-long red transient, however, is not the only evidence for the 
lanthanide-rich ejecta.
Other indications include the temperature plateau at 2500~K after $>5$~days 
\cite{drout17}, near-infrared spectral features 
\cite{pian17,chornock17,wu19,kasliwal22} and the late-time bolometric 
light curve \cite{kasliwal17,rosswog18a,villar18}.
Finally, there are new indications of features of doubly-ionized La and Ce 
in the early spectrum of AT2017gfo\cite{domoto22}.
It is worth noting that the early blue emission was detected with broad 
spectral features indicating rapid photospheric expansion with the 
velocities $>0.3$~c\cite{shappee17,chornock17}.
At the same time, the redder, lanthanide-rich component had more narrow 
features, consistent with lower expansion velocities. 
Interpreting the data on the basis of 1D spherically-symmetric models,
one finds the lanthanide-rich ejecta to be relatively massive ($>0.04$ \msun) 
and slow ($\sim0.1$~c) \cite{cowperthwaite17,villar17,drout17,perego17}.
Models that incorporate asphericity, however, allow to explain the same 
signal with smaller masses of lanthanide-rich 
ejecta \cite{wollaeger18,kawaguchi18,korobkin21} and faster expansion. 
The presence of lanthanide features in the early emission is consistent 
with small amounts of rapidly expanding dynamical ejecta \cite{kawaguchi18}.
 There is no consensus yet about the number of ejecta components, some argue that models need to include at least three ejecta  components \cite{cowperthwaite17,villar17},
while others argue that even a single composition component
can explain the basic properties \cite{waxman18}.

In the mid-IR spectrum of AT2017gfo, observations with Spit\-zer telescope 
showed strong emission at 43 days after the NS merger event GW170817 
at 4.5~$\mu$m, but nothing at 3.6~$\mu$m~\cite{villar18}.
This could indicate the existence of a strong spectral line around 
his region, which could be explained with an emission by W or Se 
ions \cite{hotokezaka22}. Features from late-time Spitzer observations 
are consistent with being produced by the decay
of the heaviest isotopes with a half-life around
$\sim 14$ days (e.g. $^{225}$Ra) \cite{wu19,kasliwal22}, but
could potentially also be explained by a composition with only light 
$r$-process elements \cite{hotokezaka20a}.

Another important class of observational constraints on the $r$-process 
production comes from  metal-poor galactic halo stars.
It has long been established that the so-called ``universal $r$-process'' 
spanning the region between Ba and Ir in metal-poor stars and in solar 
$r$-process residuals is essentially identical, within the measurement 
uncertainty \cite{sneden08,cowan21}.
This means that the primary $r$-process site responsible for the majority of the heavy r-process, must do it in a manner which respects this observational pattern.
If mergers are responsible for the $r$-process, this excludes certain channels such as, for example, winds from a hypermassive merger remnant, or shock dynamic ejecta, where electron fraction is raised and the $r$-process pattern becomes strongly sensitive to the hydrodynamics conditions.

We do not have accurate measurements of the $r$-process in GW170817, but we can compare the estimated fraction of the lanthanides to what is observed in $r$-enhanced halo stars. 
Assuming that neutron star mergers are responsible for the majority of 
the $r$-process, the estimates of the lanthanide fraction synthesized 
in such a merger yield $\log X_{\rm La}\sim-2\pm0.5$~dex, while for the 
$r$-enhanced halo stars this fraction is significantly higher: 
$\log X_{\rm La}\sim-1.5$~dex \cite{ji19}.
The general consensus about kilonova GW170817 is that the total fraction of 
synthesized lanthanides should be $\log X_{\rm La} = -2.2\pm 0.5$, which 
is therefore in tension with the $r$-enhanced halo stars \cite{ji19}.
This may mean that GW170817 was not a typical merger event, and lanthanides 
were underproduced.
Another possible way to resolve this discrepancy is the directional 
dependence in the neutron-to-proton ratio in the ejecta, which may not 
necessarily mix very well in the galactic halo.
 It is also possible that the range of $X_{\rm La}$ produced in 
neutron-star mergers mergers is broader, or that another source was 
contributing to the $r$-process at low metallicities. 
As a result, different halo stars may have been enriched by
different parts of the ejecta from a single merger.
This is an additional factor increasing variability in the observed 
lanthanide fraction within the population of halo stars.

\begin{figure*}[htp!]
  \centering
  \includegraphics[width=0.99\textwidth]{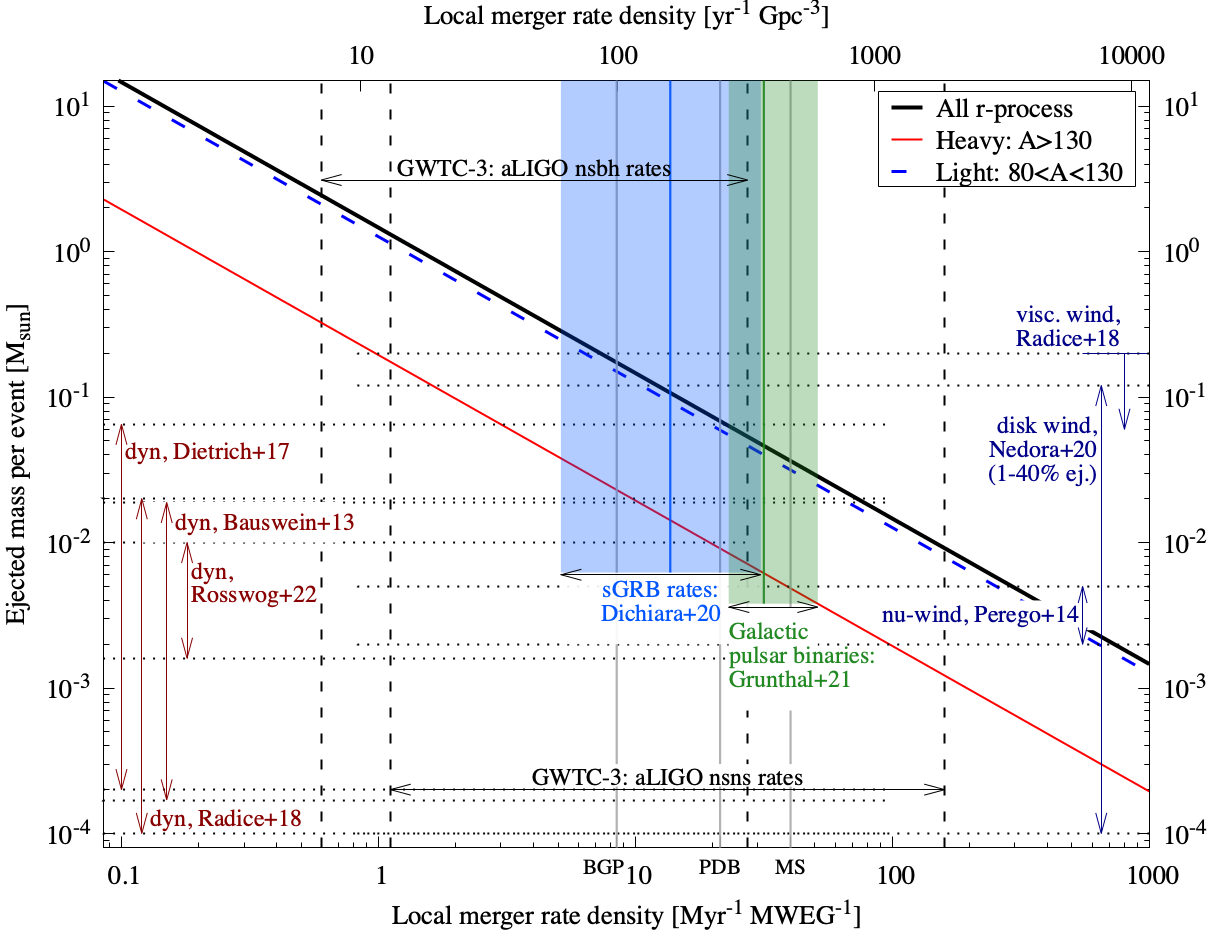}
  \caption{The figure shows how much $r$-process needs to be ejected 
  per event (diagonal lines from upper left to lower right) to explain
  the observed $r$-process material (update of Fig.2 from 
  \citeref{rosswog17a}). In addition, various rate estimates and ejecta
  masses from simulations are shown.}  The horizontal axis 
  shows the local merger rate density using different units on the 
  top and bottom axes.
  The units on the top axis are events per year in Gpc$^3$, while 
  the units on the bottom are events per millon years, per Milky 
  Way Equivalent Galaxy (MWEG).
  We use the conversion factor of 
  ${1\ {\rm Myr}^{-1}{\rm MWEG}^{-1} = 11.7\ {\rm yr}^{-1}{\rm Gpc}^{-1}}$, 
  corresponding to $1.17\times10^7$ MWEGs in 
  Gpc$^3$ \cite{kopparapu08,mandel22}.
  The vertical axis shows the ranges of ejecta masses produced via 
  different ejection channels, or, respectively, needed to produce 
  ranges of observed $r$-process material (top-left to bottom-right 
  lines). Also shown are various rate estimates, including LVC estimates 
  for nsns and nsbh binaries as well as rates based on short 
  GRBs\cite{dichiara20} and Galactic pulsar binaries\cite{grunthal21}. 
  The labels BGP, PDB and MS on the horizontal axis refer to different 
  statistical models used to process the data in Gravitational Wave 
  Transient Catalog 3 (GWTC-3) in \citeref{abbott21}:
  BGP refers to ``binned Gaussian processes'',  PDB refers to 
  ``power law + dip + break'', and MS refers to the 
  ``power law + peak multi-source'' model of \citeref{abbott21}. 
  The grey solid vertical lines indicate the expectation values of these models.
  Ranges of masses of dynamical ejecta, indicated on the left, 
  are taken from~\citeref{bauswein13a, dietrich17, radice18a, rosswog22b}.
  Disk wind mass range on the right is computed by taking $1-40$~\% of 
  the torus mass calculated in \citeref{nedora19}.
  Neutrino-driven wind mass estimate is taken from \citeref{perego14b}, 
  while upper limit on the mass of viscously ejected wind is 
  from \citeref{radice18b}.
  \label{fig:rates}
\end{figure*}

In Fig.~\ref{fig:rates} we provide an update
of our Fig.~2 from \citeref{rosswog17a}. 
If we take the solar-system $r$-process pattern \cite{arnould07} as
representative, assume a baryonic mass of the Milky Way of 
$6\times 10^{11}$ \Msun and assume $\approx 10^{10}$ yrs for the age 
of the Galaxy, we find a rough $r$-process production rate of 
$\approx 1.6 \times 10^{-6}$ \Msun yr$^{-1}$ for all $r$-process and
$\approx 2.5 \times 10^{-7}$ \Msun yr$^{-1}$ as production rate 
for the $r$-process elements heavier than $A>130$. These production 
rates must be the product of the average event rate and the average 
ejecta mass (in the considered $A$-range). For a given average event 
rate density (on the x-axis; in events per Myr$^{-1}$ and ``Milky Way 
equivalent Galaxy'' (MWEG) \cite{abadie10} on the bottom and in events 
per year and Gpc$^{-3}$ on the top), the average ejecta mass required 
to produce all of the Galactic $r$-process is shown as the black solid
line from the upper left to lower right. The corresponding lines for 
heavy  ($A>130$) and light $r$-process ($80<A \le 130$) are shown as 
solid red and dashed blue lines, respectively. The estimated neutron star merger 
and neutron star black hole merger event rates from the LIGO/Virgo 
collaboration \cite{abbott21} are also shown as well as short GRB 
rates \cite{dichiara20} and rate estimates based on Galactic pulsar 
binaries \cite{grunthal21}. In addition, we overplot ejecta masses 
from various simulation. As an example, if the average neutron star 
merger rate were $2 \times 10^{-5}$ yr$^{-1}$ MWEG$^{-1}$,
a merger would need to eject about $10^{-2}$ \Msun of 
heavy $r$-process event in order to account for the heavy $r$-content
in the Galaxy.

\section{Directions for the future}
As outlined above, a lot of theoretical groundwork has been 
laid in the last two decades with a major confirmation of
the overall big picture (and a lot of inspiration) coming from the first 
detected neutron star merger GW170817/AT2017gfo. However, many 
aspects have so far remained un- or at least insufficiently explored.
We give here a brief overview of possible future directions
and for some of these explorations the nuclear heating library 
described in Appendix~\ref{app:heating} may be useful.

The $r$-process site has remained a mystery
for so many decades, because it involves major challenges
on both the nuclear and the astrophysical side. On the nuclear 
physics side, properties of extremely neutron-rich nuclei
(close to the neutron drip line) are needed which include 
nuclear masses, $\beta$-decay half-lives, neutron capture 
and fission rates.
On the astrophysical side, all of the promising 
productions sites involve extremely strong gravitational
and likely magnetic fields and densities and temperatures
far beyond anything else that exists  in the post-Big Bang
Universe. In what follows below, we focus on the 
astrophysical side, for a discussion of the involved 
nuclear physics we refer the interested reader to 
the excellent recent review of \citeref{cowan21}.

\subsection{The scale challenge}
For GW170817, gravitational waves and, much later, a firework of
electromagnetic emission, including a kilonova,
was observed. The gravitational waves are 
shaped by matter at supra-nuclear densities,
moving at a substantial fraction of the speed 
of light in a strongly curved, dynamical spacetime. 
Close to merger, the typical time scales are $\sim$milliseconds,
and the typical length scales $\sim 10$~km. 
After the merger, in most cases, a close-to-collapse,
neutron star-like object with densities close to 
$\sim 10^{15}\ \gcc$ and temperatures
substantially exceeding $10^{11}$ K \cite{perego19a} forms, which 
emits kHz gravitational waves and radiates neutrinos 
at luminosities exceeding 
the solar luminosity by a whopping factor of $\sim 10^{20}$.

The kilonova emission, in contrast, peaks at time scales
of $\sim 1$ week, when the ejecta have expanded to $\sim10^{10}$~km, 
and the densities have dropped to $\sim 10^{-14}\ \gcc$ 
(see, e.g. Fig.~1 in \citeref{rosswog14a}). The physics 
that determines this stage is the heating by radioactive decays,
the thermaliztion of the decay products, atomic line opacities and, 
at late stages, NLTE effects in the rapidly expanding gas cloud. 
So between the observed signatures,
gravitational waves and kilonova, there is a huge gap in scales,
physics and observations and also in our understanding
of what is actually happening. 
This gap is so far usually bridged my making strongly 
simplifying assumptions such as homology and/or spherical 
symmetry, but the real picture is likely substantially more 
complicated, for example geometric effects, nuclear heating
etc. can still impact the evolution significantly 
\cite{rosswog14a,korobkin21,wu22}.
In other words, there is 
plenty of room for interesting new explorations.

\subsection{3D geometry of ejecta components}

\subsubsection{Different ejection channels and their possible interactions}
\label{sec:ejecta_interactions}
The ``easiest'' to handle ejection channel are dynamical ejecta
that are purely hydrodynamically launched. But even this, ``easy'', 
task requires at least 3D full GR hydrodynamic simulations of tens of
ms that include a temperature-dependent, nuclear equation of state
and a reliable treatment of weak interactions/neutrinos. 
Other effects such as magnetic fields and/or viscous dissipation
can likely be ignored in these early stages, but will become 
important later. Wind
ejecta simulations need to be run for substantially longer time 
scales ($\sim 100$ ms), and, due to the longer time scales, need
a fairly accurate neutrino transport and should ideally also 
include the  evolution of the magnetic field. The same physics 
ingredients, but even longer time scales ($\sim 1-10$ s) are needed 
for the torus unbinding. In 3D, this can to date only be done
for black hole torus systems, where the spacetime can be 
considered as stationary (i.e. Kerr) and the numerical time
step is set by the sound speed in the torus~\cite{darbha21,stewart22}
(rather than by the speed of light).
While current explorations starting from (non-selfgravitating)
equilibrium tori around Kerr black holes are important steps
for sharpening our understanding, it remains to be seen
how realistic they really are, given that the tori formed in 
violent mergers of neutron stars and black holes are likely
very far from being in equilibrium.
For a central neutron star, the space-time 
has to be dynamically  evolved which substantially 
increases the computational costs. To date, this 
challenge can only be met with 2D simulations \cite{fujibayashi22}.

An interesting, but so far hardly explored question
is the interaction between different ejecta components.
This could emerge, for example, if tidal dynamical ejecta
are launched first, but faster ``shock'' ejecta produced
by quasi-radial oscillations of the central remnant
catch up with them \cite{combi22}. If occurring close enough, the resulting
shock could produce large enough temperatures to raise the
electron fraction, otherwise at least a mixing between 
higher- and low-$Y_e$ material would occur. This could lead
to electron fraction distributions that are far more 
complicated than what is usually assumed in kilonova models.

Another interesting interaction could occur between matter
that is nearly unbound, but falls back after some time towards
the central source. This matter could interact with and potentially
bury late-time disk ejecta.

Virtually guaranteed interactions, of course, 
come from the relativistic, jetted
outflows that are needed to produce GRBs \cite{meszaros06,nakar07,lee07,kumar15}.
They are usually assumed to go along with
black hole formation, but a recent study
\cite{moesta20} finds that they could also 
emerge with a central neutron star being present\footnote{
The basic picture of highly magnetized neutron star-like
objects producing GRBs had been outlined already earlier,
see \citeref{usov92}}.
The interaction of jets with the other ejecta 
components may substantially impact the 
electromagnetic emission and detectability
\cite{nagakura14,murguia16,kasliwal17,lazzati17,gottlieb18,
lyman18,mooley18,nativi21a,nativi21b,murguia21}. For example,
the neutrino-driven wind simulations of \citeref{perego14b}
show low $Y_e$ material at the leading edge of the wind 
ejecta which ``shield'' inner, higher $Y_e$ ejecta.
Even very small ($\gtrsim10^{-4}$) mass fractions of lanthanides 
or actinides can be very effective in obscuring a blue kilonova 
component seen behind an ejecta component which contains lanthanides.
The latter acts as a ``curtain'' that efficiently occludes the 
visible light emitted at deeper layers 
\cite{kasen15a,wollaeger18a}. 
A jet plowing through these wind ejecta would punch a 
``hole'' into this lanthanide curtain thus allowing
bluer radiation to escape from the polar region 
\cite{nativi21a}.

\subsubsection{Multidimensional radiative transfer effects}
Several studies have begun to explore  multi-dimensional effects 
in the kilonova signals 
\cite{rosswog14a,wollaeger18,kawaguchi18,korobkin21,heinzel21} 
and found that they can have a significant influence, controlling 
an order of magnitude of the apparent kilonova luminosity. This is dramatically 
different from the initial estimates, which argued that the directional variability 
can be reduced to the projected area of the photosphere along the line of sight, 
and therefore should only affect kilonova luminosity within about a factor of 
two~\cite{grossman14a,darbha20}. While this may be a result of the used 
gray-opacity approximation, more advanced radiative transfer 
simulations with detailed, density-dependent opacities, 
reveal strong density-driven effects on light curves and
spectra~\cite{korobkin21}. 
It will be up to numerical simulations to clarify 
which range of density profiles is actually realized 
in the different ejection channels or by their
interaction.

Many numerical simulations predict  aspherical, approximately 
axisymmetric shapes of the ejecta, with a gradual change in the 
neutron-to-proton ratio as a function of the polar angle.
In the emerging picture that  the ``red'' lanthanide-rich ejecta 
have approximately toroidal shape, while the polar regions are 
occupied by  lighter $r$-process, ``blue'' ejecta.
But for mass ratios $q$ deviating substantially from unity, 
the tidal ejecta can form spiral arms that break axisymmetry, under
certain conditions dynamical ejecta can produce ``lanthanide pockets''
\cite{combi22} and long-term simulations of post-merger disks
 show radially dependent composition profiles that
are relics of the accretion history \cite{li21}. Such
effects are usually neglected in kilonova model explorations 
(with some notable exceptions\cite{bulla21}). On the other hand, 
the dynamic ejecta that result from shocks possess sufficiently 
high entropy to form an almost-spherical remnant after its 
expansion to homology \cite{rosswog22b}.

Another way in which the shape of the ejecta can play a tremendous 
role is via the above described ``lanthanide curtaining''.
This ``lanthanide curtaining'' could manifest itself as an 
``opening angle'' around the kilonova pole, being the maximum 
angle at which the blue emission is still visible.
Perhaps in the future, two categories of kilonovae will be 
identified: some with  rapidly-rising, short-lived blue transients, 
and others without, where ``lanthanide curtaining'' is operating. 

In a multi-component, multi-dimensional setting, several other effects can
manifest themselves and must be considered when interpreting kilonovae signals,
such as e.g. flux reprocessing \cite{kawaguchi18}, or the orientation dependence
of the width of the spectral features such as P~Cygni features \cite{korobkin21}.
Higher expansion velocities across the line of sight smear out even 
broad features and present significant stumbling blocks for identifying
elements.

The shape of the ejecta right after their launch  by the central 
engine can differ significantly from their shape at the kilonova 
epoch \cite{rosswog14a}.
The $r$-process occurring in the ejecta lasts for about one
second and can generate a sizeable fraction of the total 
kinetic energy of expansion and thus modify the distribution 
of ejecta mass in the velocity space.
The peak nuclear energy injection happens at around one second, 
and depends on the expansion velocity and in
particular the neutron richness (see Appendix~\ref{app:heating}).
On the second and subsecond timescales, variations in the 
nucleosynthesis in neighboring parcels of expanding fluid can 
create fluid motions and possibly internal shocks.
More work needs to be done to understand the extent of this 
reshaping and related possibilities of internal shocks due to 
inhomogeneities in nuclear heating.
The nuclear heating library that is provided with this
study  covers the range of timescales
where the ejecta reshaping is happening, and could help
understanding such effects.

Much later, at the kilonova epoch, the ejecta is in the
state of free homologous expansion, when different fluid 
parcels are out of sonic contact with each other and the
changes in its morphology are completely negligible.

\subsection{NLTE effects and spectra at late times}

At the time of writing, despite considerable efforts, 
no simulated spectra have been produced for the late-time 
epoch ($>7$ days) that match AT2017gfo well enough to 
deduce the presence, let alone abundances, of individual 
elements (evidence for the element Sr in early spectra 
is still inconclusive).
In particular, the anomalously bright visible bands 
at late epoch are in tension with all current models 
that use detailed opacities~\cite{tanvir17}.

It is highly likely that the assumption of local 
thermodynamic equilibrium (LTE), employed in opacity 
calculations and in radiative transfer simulations, is 
the one to blame for these discrepancies between theory 
and observation.
Relaxing this assumption to produce so-called ``NLTE'' 
simulations of opacity and radiative transfer, is not 
possible at present time.
All currently used radiative transfer models use the 
assumption of LTE, which means that the populations of 
ionized and excited states are calculated using a 
single temperature, by solving the Saha-Boltzmann equation.
Fully removing the assumption of LTE in atomic population 
calculations is a daunting computational task for the heavy 
elements produced in the $r$-process, due to the colossal
number of atomic energy levels involved, and the need to 
invert a matrix with the rank equal to the number of 
those levels, associated with a system of coupled rate 
equations~\cite{fontes20}.

In spite of these difficulties, work has been started on 
the nebular phase of kilonovae, where NLTE effects play 
the dominant role~\cite{hotokezaka21,pognan22a,pognan22b}.
Elements Se ($Z = 34$) and W ($Z = 74$) have been implicated 
in producing a peculiar line feature at 4.5~$\mu$m at 
the nebular phase.~\cite{hotokezaka22}.

Estimates of the rates of kinetic processes in the nebular 
phase indicate that the so-called coronal approximation is 
applicable at late epochs~\cite{pognan22a}.
Because the ejecta do not have a photosphere in nebular 
phase, its cooling is no longer efficient, and the rates 
of individual kinetic processes must be considered.
In the coronal approximation, the radiative field is considered negligible in
determining the excitation and ionization structure, and therefore the material
emissivity can be considered as only dependent on the plasma density
and effective electron temperature.
Moreover, electronic excitations and ionizations are dominated 
by collisions with fast electron populations generated by 
radioactivity, rather than by collisions with thermal Maxwellian electrons that are strongly subdominant.
The energy budget in the electron population is balanced by the influx 
of new high-energy $\beta$-particles produced by radioactivity, and by
inefficient NLTE cooling via collisions with ions. The NLTE spectrum
directly reflects the $\beta$-radioactivity in the ejecta, so the nebular 
emission can be thought of as an $r$-process radioactivity, cascaded 
down to the visible and infrared range. Simulating realistic spectra in 
such regime may require not only a NLTE treatment, but also 
detailed electron transport models which go beyond the single kinetic 
temperature~\cite{hotokezaka21}.

\subsection{Electron thermalization}

Thermalization of high-energy decay radiation directly affects the luminosity of a kilonova.
Unlike supernovae where $\gamma$-radiation is the dominant source of heating, the $r$-process powered kilonovae are heated predominantly by $\beta$-particles.
While under certain conditions of neutron richness, specific nuclear mass models 
predict that fission products dominate the nuclear heating \cite{zhu18}, for 
example in the case of the fission of $^{254}$Cf, in the majority of interesting cases
the nuclear heating on kilonova time scales is due to
the energy released in $\beta$-decays 
\cite{hotokezaka17a}.
A few simple, but powerful prescriptions were derived for
the thermalization efficiency \cite{barnes16a,hotokezaka20}. These prescriptions,
however, assume a rather simple homogeneous background.
As explained above, a population of such fast electrons  determines the nebular emission \cite{pognan22b}.
Current challenges include the modeling of
populations of such high-energy electrons.
The latter depends on the nuclear physics and, in addition, on the state of the involved 
magnetic fields.
If magnetic fields in neutron stars are on the order of $10^{12}$~G, then conservation of magnetic flux confines the electrons to Larmor radii that are about $10^{-4}$ of the ejecta radius at 100~d \cite{barnes16a,waxman19,hotokezaka21}.
A comprehensive code with full electron transport is required to better understand the thermalization process.

In the future, full electron transport codes will be needed to validate the existing approximations and to explore the extent to which kilonovae are sensitive to it.
In the provided heating library, roughly 25$\%$ of the energy can be considered to come from $\beta$-particles (with the rest of the energy being lost to $\gamma$-photons and neutrinos at late times).
An average energy of a $\beta$-particle is $\sim0.5\pm0.2$~MeV\cite{barnes16a}; these numbers, along with the heating rates, can be used as an input to  electron transport codes for corresponding sensitivity studies.

\section{Summary}
One of the arguably most important consequences of a neutron star merger
is that it ejects a fraction of its mass as neutron-rich matter into space. This
matter experiences rapid neutron capture nucleosynthesis and the decay of the
freshly synthesized, unstable nuclei powers a thermal, relatively iso\-tropic
electromagnetic transient. Such transients provide precious {\em additional}
information that is not contained in the gravitational wave signal which delivers insights
into both the extreme physics at work during a merger as well as into the astronomical
environment where it occurs.

Just a few months prior to the start of the fourth LIGO-Virgo-KAGRA observation
run (``O4''), we summarized here our current understanding of the various ejection
channels, the expected ejecta properties and how this translates into potentially 
observable thermonuclear transients. We also stated important issues where
no agreement has been reached yet and we discussed possible ways forward, into
so far unexplored scientific territory. And we hope, of course, that soon 
more light will be shed on this exciting topic by future multi-messenger 
observations.

\section*{Acknowledgements}
SR would like to thank Eli Waxman for the invitation to a very inspiring
workshop on the GW-EM connection at the Weizmann Institute. Thanks also to all the
participants for creating a constructive brainstorming atmosphere.
OK is grateful to Ryan T. Wollaeger, Jonah M. Miller and 
Chris L. Fryer for helpful comments and spirited discussions.
We are further grateful to Jonah Miller, Udi Nakar, Quantin Pognan, 
David Radice, Daniel Siegel and the two anonymous referees
for their careful reading of a first version of this manuscript and 
for their insightful comments.

SR has been supported by the Swedish Research Council (VR) under 
grant number 2020-05044, by the research environment grant
``Gravitational Radiation and Electromagnetic Astrophysical
Transients'' (GREAT) funded by the Swedish Research Council (VR) 
under Dnr 2016-06012, by the Knut and Alice Wallenberg Foundation
under grant Dnr. KAW 2019.0112, by the Deutsche 
Forschungsgemeinschaft (DFG, German Research Foundation) 
under Germany's Excellence Strategy -EXC 2121-
"Quantum Universe" - 390833306 and by the European Research Council (ERC) Advanced Grant
INSPIRATION under the European Union's Horizon 2020 research and innovation programme
(Grant agreement No. 101053985).
SR's calculations were performed 
on the facilities of the North-German Supercomputing Alliance (HLRN), 
on the resources provided by the Swedish National Infrastructure for 
Computing (SNIC) in Link\"oping, partially funded by the Swedish Research 
Council through Grant Agreement no. 2016-07213, and at the SUNRISE HPC 
facility supported by the Technical Division at the Department of 
Physics, Stockholm University. Special thanks go to Holger Motzkau 
and Mikica Kocic for their excellent support in upgrading and 
maintaining SUNRISE.
This work was supported by the US Department of Energy through the Los Alamos National Laboratory. Los Alamos National Laboratory is operated by Triad National Security, LLC, for the National Nuclear Security Administration of U.S.\ Department of Energy (Contract No.\ 89233218-CNA000001).
Research presented in this article was supported by the Laboratory Directed Research and Development program of Los Alamos National Laboratory under pro\-ject number 20200145ER.
All LANL calculations were performed on LANL Institutional Computing resources.

\section*{Conflict of Interest}

The authors declare no conflict of interest.
Approved for public release with designated number LA-UR-22-25522; distribution is unlimited.




\appendix  

\section{Appendix: Publicly available nuclear heating information}
\label{app:heating}
Here we provide a concise description of a publicly available heating rate library that is based on the Winnet network \cite{winteler12,winteler12b}. Winnet itself  is based on the BasNet network \cite{thielemann11} and it contains 5831 isotopes between the valley of stability and the neutron drip line, starting with nucleons and reaching up 
to $Z=111$. The implemented reaction rates are from the compilation of
\citeref{rauscher00} and based on the Finite Range Droplet Model (FRDM) \cite{moeller95}.
The electron-/positron capture and $\beta$-decay rates are taken from \citeref{fuller82} and \citeref{langanke01}, the neutron capture and neutron-induced fission rates are due to \citeref{panov10} and the $\beta$-delayed fission probabilities from \citeref{panov05} are implemented. Below, we provide a library of heating rates as direct output from Winnet and we further provida a 
simple fit formula which may be more convenient for the implementation of nuclear heating  into hydrodynamic simulations.

\subsection*{Heating library}
We produced a heating rate library that is based on paramet\-rized
trajectories that cover a broad range of conditions relevant for
neutron star merger ejecta. The setup of the trajectories is the same
as described in Sec. 2.2 and 2.3 of our earlier work \cite{rosswog17a}. For the mass 
that is used to set up the initial trajectory, we adopt $m_{\rm ej}= 0.05$ \msun,
close to the estimate for GW170817
\cite{kasen17,cowperthwaite17,evans17,villar17,kasliwal17,tanvir17,rosswog18a}. 
The energy output is in units of erg/(g s) (i.e. independent of the ejecta
mass) and our value for $m_{\rm ej}$ only impacts the starting point of the trajectory, to 
which the heating rate is hardly sensitive. Another quantity that enters the calculation
of the trajectory is the initial entropy $s_0$. The nucleosynthesis (and therefore heating rate)
at low values of $Y_e$ is hardly impacted by  
the initial entropy \cite{freiburghaus99a},
therefore, for tidal ejecta the heating rate is insensitive to the chosen entropy value. For neutrino-driven winds
the entropy distribution is strongly peaked at $s\approx 15$ $k_{\rm B}$/nucleon \cite{perego14b,just15,fernandez15} and we therefore choose this value for $s_0$ to determine the
trajectory. Within the heating rate library, we vary the electron fraction $Y_e$ between 0.05 and 0.5
in steps of $\Delta Y_e= 0.05$ and 
we explore velocity values of $[0.05c, 0.10c, 0.2c, 0.3c, 0.4c, 0.5c]$.\\
The corresponding heating histories, time in days and ``naked'' heating rate $d\epsilon/dt$ in erg/(g s) (i.e.
no heating efficiency applied) and the resultung abundances can be downloaded at\\
\url{http://compact-merger.astro.su.se/downloads.html}.

\subsection*{Fit formula}
The following 13-parameter expression provides an acceptable (within 2$\%$) fit for the simulated heating rates:
\begin{align}
\frac{d\varepsilon}{dt} &= \dot{\varepsilon}_0
\left(\frac12 - \frac1\pi\arctan\left[\frac{t-t_0}{\sigma}\right]\right)^\alpha
\left(\frac12 + \frac1\pi\arctan\left[\frac{t-t_1}{\sigma_1}\right]\right)^{\alpha_1} \nonumber \\
&+ C_1 e^{-t/\tau_1} 
 + C_2 e^{-t/\tau_2}
 + C_3 e^{-t/\tau_3}
\end{align}

Here, the first term in parentheses models a constant heating rate with a break to a power-law decay at $t_0$, with a power-law index $\alpha$ and a transition width $\sigma$.
The next term models a subsecond dynamics of the heating rate by a power-law raise with an index $\alpha_1$, threshold $t_1$ and a transition width $\sigma_1$. 
The last two terms are needed to model the cases when individual isotopes dominate the power-law decay heating rates.\\
What we have just described is the raw nuclear energy generation rate.
To obtain the energy input into the macronova/kilonova, some prescription for the
thermalization efficiency needs to be applied, see e.g. \citeref{hotokezaka20}.




\begin{figure*}[htp!]
  \begin{tabular}{cc}
  \includegraphics[width=0.5\linewidth]{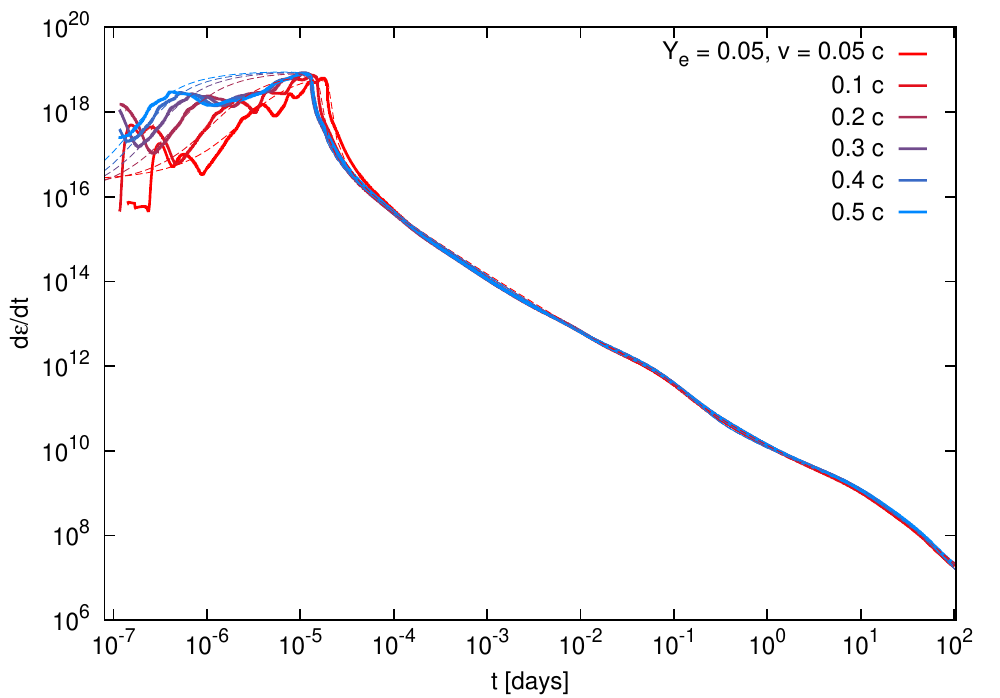} &
  \includegraphics[width=0.5\linewidth]{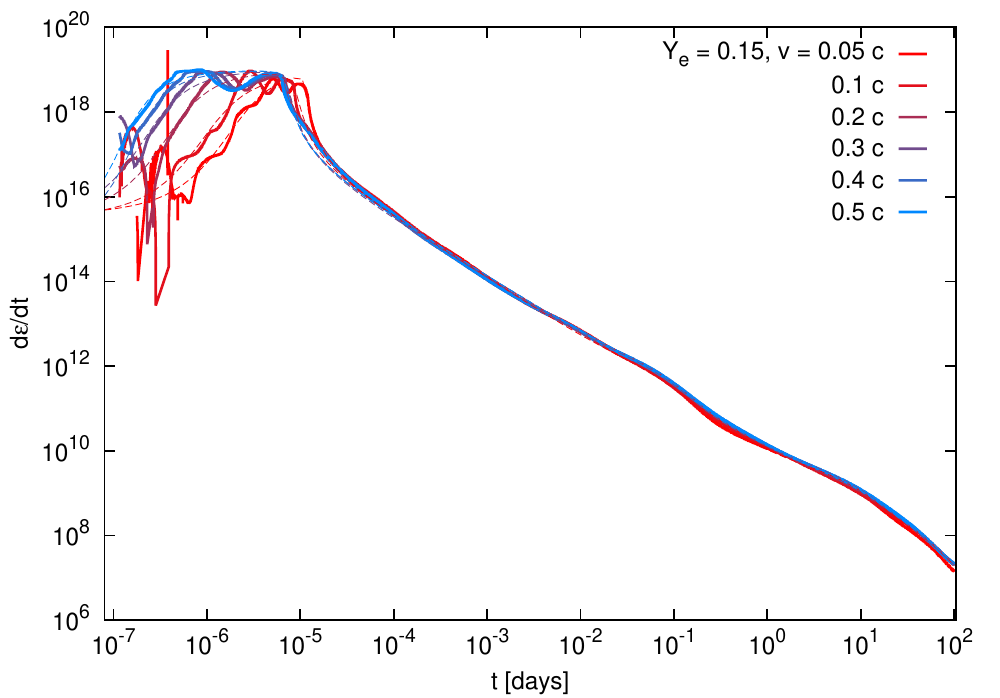}
  \\
  \includegraphics[width=0.5\linewidth]{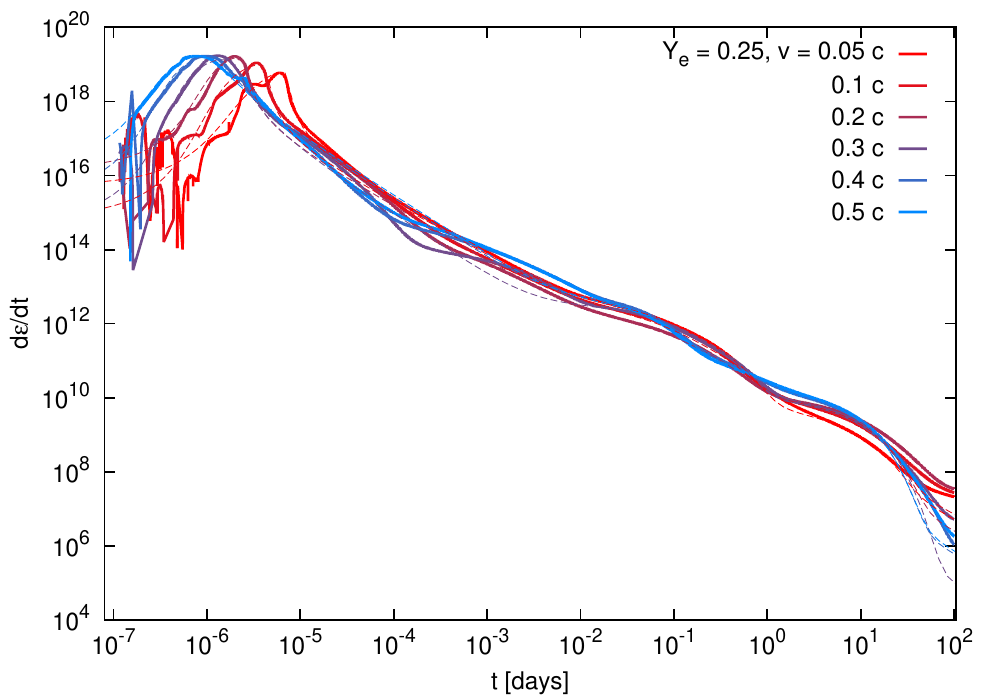} &
  \includegraphics[width=0.5\linewidth]{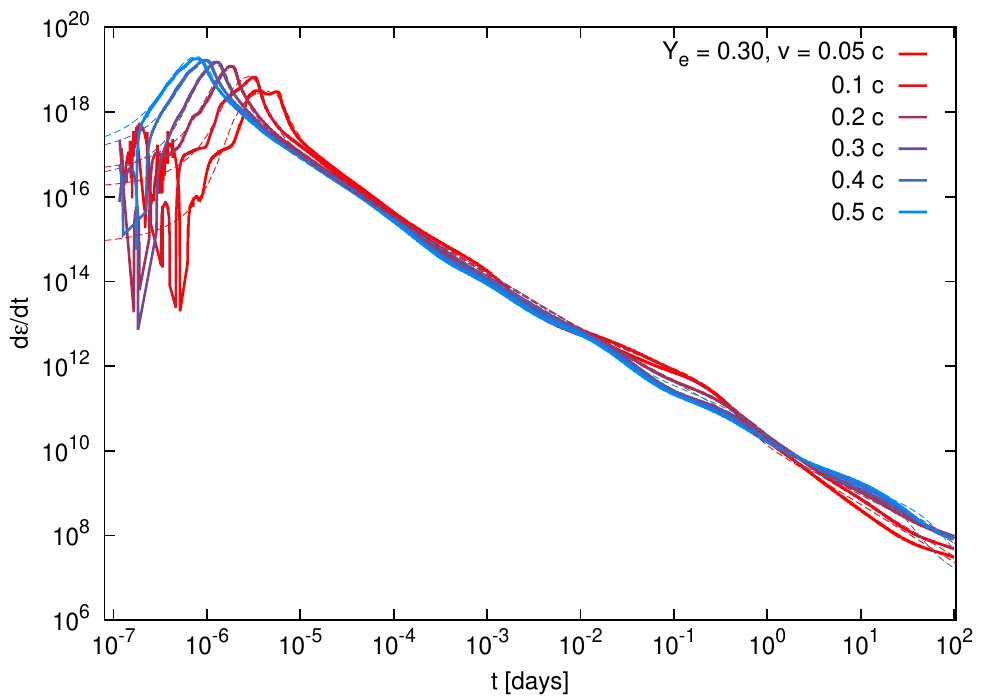}
  \\
  \includegraphics[width=0.5\linewidth]{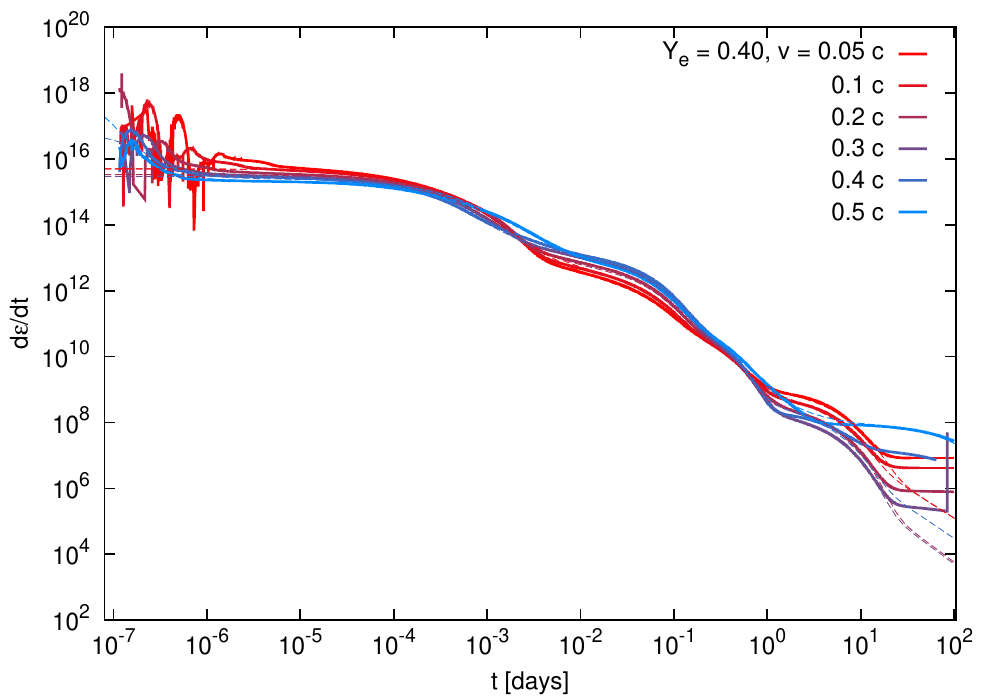} &
  \includegraphics[width=0.5\linewidth]{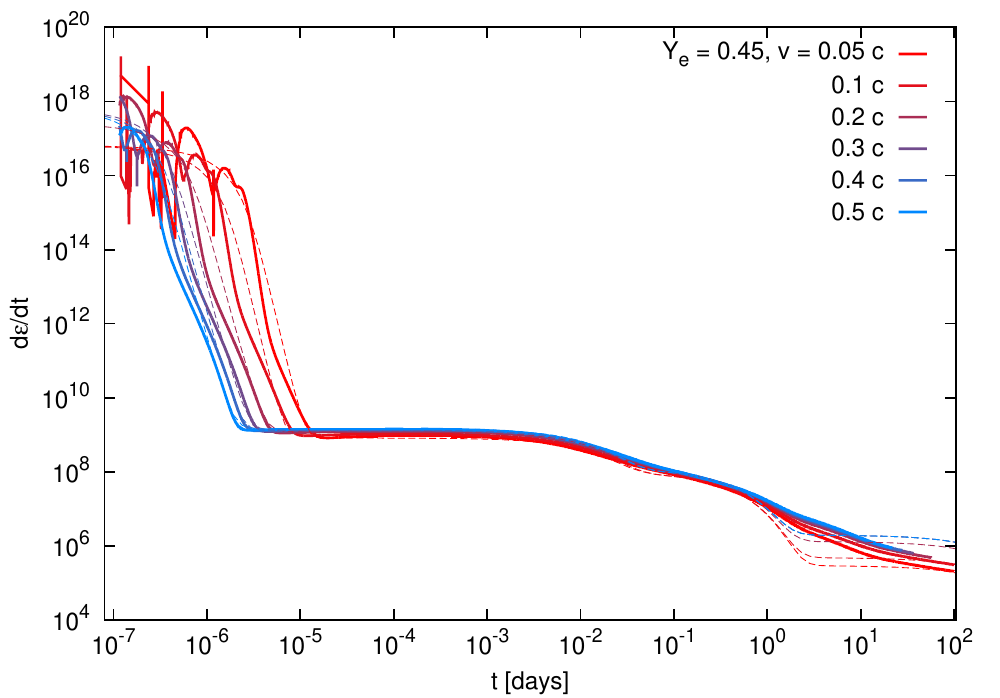}
  \end{tabular}
  \caption{Example fits of the radioactive heating rate.
  Solid lines represent numerical rates computed with the nucleosynthesis network,
  and dashed lines are the fits.}
  \label{fig:boat1}
\end{figure*}

\begin{table*}
 \caption{Fit coefficients}
  \begin{tabular}[htbp]{l|lcccccc|lcccccc}
    \hline
    par.$\backslash v_{\rm ej}$
    & $Y_e$ & 0.05c& 0.1c & 0.2c & 0.3c & 0.4c & 0.5c 
    & $Y_e$ & 0.05c& 0.1c & 0.2c & 0.3c & 0.4c & 0.5c 
    \\
    \hline
$\dot\epsilon_0\times10^{18}$ & 0.05  & 10.0 & 10.0 & 10.0 & 10.0 & 10.0 & 10.0  & 0.1 &10.0  & 10.0  & 11.0  & 11.0  & 11.0  & 11.0 \\
$\alpha$                      &       & 1.37 & 1.38 & 1.41 & 1.41 & 1.41 & 1.41  &     &1.41  & 1.38  & 1.37  & 1.37  & 1.37  & 1.37  \\
$t_0$ (s)                     &       & 1.80 & 1.40 & 1.20 & 1.20 & 1.20 & 1.20  &     &1.40  & 1.00  & 0.85  & 0.85  & 0.85  & 0.85  \\
$\sigma$ (s)                  &       & 0.08 & 0.08 & 0.095& 0.095& 0.095& 0.095 &     &0.10  & 0.08  & 0.07  & 0.07  & 0.07  & 0.07  \\
$\alpha_1$                    &       & 7.50 & 7.50 & 7.50 & 7.50 & 7.50 & 7.50  &     &9.00  & 9.00  & 7.50  & 7.50  & 7.00  & 7.00  \\
$t_1$ (s)                     &       & 0.040& 0.025& 0.014& 0.010& 0.008& 0.006 &     &0.040 & 0.035 & 0.020 & 0.012 & 0.010 & 0.008 \\
$\sigma_1$ (s)                &       & 0.250& 0.120& 0.045& 0.028& 0.020& 0.015 &     &0.250 & 0.060 & 0.035 & 0.020 & 0.016 & 0.012 \\
$\log{C_1}$                   &       & 27.2 & 27.8 & 28.2 & 28.2 & 28.2 & 28.2  &     &28.0  & 27.8  & 27.8  & 27.8  & 27.8  & 27.8  \\
$\tau_1$ ($10^3$ s)           &       & 4.07 & 4.07 & 4.07 & 4.07 & 4.07 & 4.07  &     &4.07  & 4.07  & 4.07  & 4.07  & 4.07  & 4.07  \\
$\log{C_2}$                   &       & 21.5 & 21.5 & 22.1 & 22.1 & 22.1 & 22.1  &     &22.3  & 21.5  & 21.5  & 21.8  & 21.8  & 21.8  \\
$\tau_2$ ($10^5$ s)           &       & 4.62 & 4.62 & 4.62 & 4.62 & 4.62 & 4.62  &     &4.62  & 4.62  & 4.62  & 4.62  & 4.62  & 4.62  \\
$\log{C_3}$                   &       & 19.4 & 19.8 & 20.1 & 20.1 & 20.1 & 20.1  &     &20.0  & 19.8  & 19.8  & 19.8  & 19.8  & 19.8  \\
$\tau_3$ ($10^5$ s)           &       & 18.2 & 18.2 & 18.2 & 18.2 & 18.2 & 18.2  &     &18.2  & 18.2  & 18.2  & 18.2  & 18.2  & 18.2  \\
\hline
$\dot\epsilon_0\times10^{18}$ & 0.15  & 14.0 & 10.0 & 11.0 & 11.0 & 11.0 & 11.0  & 0.2 &14.0 & 10.0 & 10.0 & 10.0 & 11.0 & 11.0 \\
$\alpha$                      &               & 1.41 & 1.38 & 1.37 & 1.37 & 1.37 & 1.37  &           &1.36 & 1.25 & 1.32 & 1.32 & 1.34 & 1.34  \\
$t_0$ (s)                     &               & 1.00 & 0.80 & 0.65 & 0.65 & 0.61 & 0.61  &           &0.85 & 0.60 & 0.45 & 0.45 & 0.45 & 0.45  \\
$\sigma$ (s)                  &               & 0.07 & 0.08 & 0.070& 0.065& 0.070& 0.070 &           &0.040& 0.030& 0.05 & 0.05 & 0.05 & 0.050 \\
$\alpha_1$                    &               & 8.00 & 8.00 & 7.50 & 7.50 & 7.00 & 7.00  &           &8.00 & 8.00 & 7.50 & 7.50 & 7.00 & 7.00  \\
$t_1$ (s)                     &               & 0.080& 0.040& 0.020& 0.012& 0.012& 0.009 &           &0.080& 0.040& 0.030& 0.018& 0.012& 0.009 \\
$\sigma_1$ (s)                &               & 0.170& 0.090& 0.035& 0.020& 0.012& 0.009 &           &0.170& 0.070& 0.035& 0.015& 0.012& 0.009 \\
$\log{C_1}$                   &               & 27.5 & 27.0 & 27.8 & 27.8 & 27.8 & 27.8  &           &28.8 & 28.1 & 27.8 & 27.8 & 27.5 & 27.5  \\
$\tau_1$ ($10^3$ s)           &               & 4.07 & 4.07 & 4.07 & 4.07 & 4.07 & 4.07  &           &4.07 & 4.07 & 4.07 & 4.07 & 4.07 & 4.07  \\
$\log{C_2}$                   &               & 22.0 & 21.5 & 21.5 & 22.0 & 21.8 & 21.8  &           &23.5 & 22.5 & 22.1 & 22.0 & 22.2 & 22.2  \\
$\tau_2$ ($10^5$ s)           &               & 4.62 & 4.62 & 4.62 & 4.62 & 4.62 & 4.62  &           &4.62 & 4.62 & 4.62 & 4.62 & 4.62 & 4.62  \\
$\log{C_3}$                   &               & 19.9 & 19.8 & 19.8 & 19.8 & 19.8 & 19.8  &           &5.9  & 9.8 & 23.5 & 23.5 & 23.5 & 23.5   \\
$\tau_3$ ($10^5$ s)           &               & 18.2 & 18.2 & 18.2 & 18.2 & 18.2 & 18.2  &           &18.2 & 18.2 & 0.62 & 0.62 & 0.62 & 0.62  \\

\hline
  \end{tabular}
\end{table*}

\begin{table*}
 \caption{Fit coefficients (continued)}
  \begin{tabular}[htbp]{l|lcccccc|lcccccc}
    \hline
    par.$\backslash v_{\rm ej}$
    & $Y_e$ & 0.05c& 0.1c & 0.2c & 0.3c & 0.4c & 0.5c 
    & $Y_e$ & 0.05c& 0.1c & 0.2c & 0.3c & 0.4c & 0.5c 
    \\
\hline
$\dot\epsilon_0\times10^{18}$ & 0.25  & 20.0 & 25.0 & 40.0 & 38.0 & 58.0 & 70.0  & 0.3 &6.1  & 18.0 & 47.1 & 47.1 & 74.8 & 74.8  \\
$\alpha$                      &               & 1.44 & 1.40 & 1.46 & 1.66 & 1.60 & 1.60  &           &1.36 & 1.33 & 1.33 & 1.33 & 1.374& 1.374 \\
$t_0$ (s)                     &               & 0.65 & 0.38 & 0.22 & 0.18 & 0.12 & 0.095 &           &0.540& 0.31 & 0.18 & 0.13& 0.095& 0.081  \\
$\sigma$ (s)                  &               & 0.05 & 0.030& 0.025& 0.045& 0.05 & 0.05  &           &0.11 & 0.04 & 0.021& 0.021& 0.017& 0.017 \\
$\alpha_1$                    &               & 8.00 & 8.00 & 5.00 & 7.50 & 7.00 & 6.50  &           &4.5  & 3.8    & 4    & 4    & 4    & 4 \\
$t_1$ (s)                     &               & 0.080& 0.060& 0.065& 0.028& 0.020& 0.015 &           &  0.14& 0.123& 0.089& 0.060& 0.045& 0.031 \\
$\sigma_1$ (s)                &               & 0.170& 0.070& 0.050& 0.025& 0.020& 0.020 &           &0.065& 0.067& 0.053& 0.032& 0.032& 0.024 \\
$\log{C_1}$                   &               & 28.5 & 28.0 & 27.5 & 28.5 & 29.2 & 29.0  &           &25.0 & 27.5 & 25.8 & 20.9 & 29.3  & 1.0  \\
$\tau_1$ ($10^3$ s)           &               & 4.77 & 4.77 & 4.77 & 4.77 & 4.07 & 4.07  &           &4.77 & 4.77 & 28.2 & 1.03& 0.613  & 1.0  \\
$\log{C_2}$                   &               & 22.0 & 22.8 & 23.0 & 23.0 & 23.5 & 23.5  &           &10.0 & 0    & 0    & 19.8 & 22.0 & 21.0  \\
$\tau_2$ ($10^5$ s)           &               & 5.62 & 5.62 & 5.62 & 5.62 & 4.62 & 4.62  &           &5.62 & 5.18 & 5.18 & 34.7 & 8.38 & 22.6  \\
$\log{C_3}$                   &               & 27.3 & 26.9 & 26.6 & 27.4 & 25.8 & 25.8  &           &27.8 & 26.9 & 18.9 & 25.4 & 24.8 & 25.8  \\
$\tau_3$ ($10^5$ s)           &               & 0.18 & 0.18 & 0.18 & 0.18 & 0.32 & 0.32  &           &0.12 & 0.18 & 50.8 & 0.18 & 0.32 & 0.32  \\
\hline
$\dot\epsilon_0\times10^{18}$ & 0.35 & 7.3    & 7   & 16.3 & 23.2 & 43.2  & 150  & 0.4 & 0.0032& 0.0032& 0.008& 0.007& 0.009& 0.015 \\
$\alpha$                      &      & 1.40 & 1.358& 1.384& 1.384& 1.384& 1.344  &     & 1.80  & 1.80  & 2.10 & 2.10 & 1.90 & 1.90  \\
$t_0$ (s)                     &      & 0.385& 0.235  & 0.1 & 0.06& 0.035& 0.025  &     & 26.0  & 26.0  & 0.4  & 0.4  & 0.12 & -20   \\
$\sigma$ (s)                  &      & 0.10 & 0.094& 0.068 & 0.05 & 0.03 & 0.01  &     & 45    & 45    & 45   & 45   & 25   & 40    \\
$\alpha_1$                    &      &   2.4   & 3.8  & 3.8 & 3.21 & 2.91 & 3.61 &     &-1.55 & -1.55 & -0.75& -0.75& -2.50 & -5.00 \\
$t_1$ (s)                     &      & 0.264  & 0.1 & 0.07& 0.055& 0.042& 0.033  &     & 1.0   & 1.0   & 1.0  & 1.0 & 0.02  & 0.01  \\
$\sigma_1$ (s)                &      & 0.075& 0.044 & 0.03 & 0.02 & 0.02& 0.014  &     &10.0  & 10.0  & 10.0 & 10.0 & 0.02  & 0.01  \\
$\log{C_1}$                   &      & 28.7  & 27.0 & 28.0 & 28.0 & 27.4 & 25.3  &     &28.5  & 29.1  & 29.5 & 30.1 & 30.4  & 29.9  \\
$\tau_1$ ($10^3$ s)           &      &   3.4  & 14.5 & 11.4 & 14.3 & 13.3 & 13.3 &     &2.52  & 2.52  & 2.52 & 2.52 & 2.52  & 2.52  \\
$\log{C_2}$                   &      & 26.2  & 14.1 & 18.8 & 19.1 & 23.8 & 19.2  &     &25.4  & 25.4  & 25.8 & 26.0 & 26.0  & 25.8  \\
$\tau_2$ ($10^5$ s)           &      &   0.15 & 4.49 & 95.0 & 95.0 & 0.95  & 146 &     &0.12  & 0.12  & 0.12 & 0.12 & 0.12  & 0.14  \\
$\log{C_3}$                   &      & 22.8  & 17.9 & 18.9 & 25.4 & 24.8 & 25.5  &     &20.6  & 20.2  & 19.8 & 19.2 & 19.5  & 18.4  \\
$\tau_3$ ($10^5$ s)           &      &   2.4  & 51.8 & 50.8 & 0.18 & 0.32 & 0.32 &     & 3.0   & 2.5   & 2.4  & 2.4  & 2.4  & 60.4  \\
\hline
$\dot\epsilon_0\times10^{18}$  & 0.45 & 0.20   & 0.20  & 0.60  & 1.50  & 1.50   & 1.50  & 0.5 &0.40  & 1.00  & 2.00   & 3.00  & 3.00  & 3.00  \\
$\alpha$                       &      & 8.00   & 8.00  & 7.00  & 7.00  & 7.00   & 7.00  &     &1.40  & 1.40  & 1.40   & 1.60  & 1.60  & 1.60  \\
$t_0$ (s)                      &      & 0.20   & 0.12  & 0.05  & 0.03  & 0.025  & 0.021 &     &0.16  & 0.08  & 0.04   & 0.02  & 0.018 & 0.016 \\
$\sigma$ (s)                   &      & 0.20   & 0.12  & 0.05  & 0.03  & 0.025  & 0.021 &     &0.03  & 0.015 & 0.007  & 0.01  & 0.009 & 0.007 \\
$\alpha_1$                     &      & -1.55  & -1.55 & -1.55 & -1.55 & -1.55  & -1.55 &     &3.00  & 3.00  & 3.00   & 3.00  & 3.00  & 3.00  \\
$t_1$ (s)                      &      &  1.0   & 1.0   & 1.0   & 1.0   & 1.0    & 1.0   &     &0.04  & 0.02  & 0.01   & 0.002 & 0.002 & 0.002 \\
$\sigma_1$ (s)                 &      & 10.0   & 10.0  & 10.0  & 10.0  & 10.0   & 10.0  &     &0.01  & 0.005 & 0.002  & $10^{-4}$  & $10^{-4}$  & $10^{-4}$  \\
$\log{C_1}$                    &      & 20.4   & 20.6  & 20.8  & 20.9  & 20.9   & 21.0  &     &29.9  & 30.1  & 30.1   & 30.2  & 30.3  & 30.3  \\
$\tau_1$ ($10^3$ s)            &      & 1.02   & 1.02  & 1.02  & 1.02  & 1.02   & 1.02  &     &0.22  & 0.22  & 0.22   & 0.22  & 0.22  & 0.22  \\
$\log{C_2}$                    &      & 18.4   & 18.4  & 18.6  & 18.6  & 18.6   & 18.6  &     &27.8  & 28.0  & 28.2   & 28.2  & 28.3  & 28.3  \\
$\tau_2$ ($10^5$ s)            &      & 0.32   & 0.32  & 0.32  & 0.32  & 0.32   & 0.32  &     &0.02  & 0.02  & 0.02   & 0.02  & 0.02  & 0.02  \\
$\log{C_3}$                    &      & 12.6   & 13.1  & 14.1  & 14.5  & 14.5   & 14.5  &     &24.3  & 24.2  & 24.0   & 24.0  & 24.0  & 23.9  \\
$\tau_3$ ($10^5$ s)            &      &  200    & 200   & 200   & 200   & 200    & 200  &     &8.76  & 8.76  & 8.76   & 8.76  & 8.76  & 8.76  \\
\hline
  \end{tabular}
\end{table*}

\medskip

\bibliographystyle{MSP}
\bibliography{refs, astro_SKR1, astro_SKR2}

\hyphenation{Post-Script Sprin-ger}
\begin{thebibliography}{100}
\providecommand{\url}[1]{\texttt{#1}}
\providecommand{\urlprefix}{URL }

\bibitem{cameron57}
A.~G.~W. Cameron,
\newblock \emph{Chalk River Rept.} \textbf{1957}, \emph{CRL-41}.

\bibitem{burbidge57}
E.~M. {Burbidge}, G.~R. {Burbidge}, W.~A. {Fowler}, F.~{Hoyle},
\newblock \emph{Reviews of Modern Physics} \textbf{1957}, \emph{29} 547.

\bibitem{woosley94}
S.~E. Woosley, J.~R. Wilson, G.~J. Mathews, R.~D. Hoffman, B.~S. Meyer,
\newblock \emph{ApJ} \textbf{1994}, \emph{433} 229.

\bibitem{takahashi94}
K.~Takahashi, J.~Witti, H.-T. Janka,
\newblock \emph{A\&A} \textbf{1994}, \emph{286} 857.

\bibitem{hoffman97}
R.~D. Hoffman, S.~E. Woosley, Y.-Z. Qian,
\newblock \emph{ApJ} \textbf{1997}, \emph{482} 951.

\bibitem{farouqi10}
K.~{Farouqi}, K.-L. {Kratz}, B.~{Pfeiffer}, T.~{Rauscher}, F.-K. {Thielemann},
  J.~W. {Truran},
\newblock \emph{ApJ} \textbf{2010}, \emph{712} 1359.

\bibitem{arcones07}
A.~{Arcones}, H.-T. {Janka}, L.~{Scheck},
\newblock \emph{A \& A} \textbf{2007}, \emph{467} 1227.

\bibitem{roberts10}
L.~F. {Roberts}, S.~E. {Woosley}, R.~D. {Hoffman},
\newblock \emph{ApJ} \textbf{2010}, \emph{722} 954.

\bibitem{fischer10}
T.~{Fischer}, S.~C. {Whitehouse}, A.~{Mezzacappa}, F.-K. {Thielemann},
  M.~{Liebend{\"o}rfer},
\newblock \emph{A \& A} \textbf{2010}, \emph{517} A80.

\bibitem{huedepohl10}
L.~{H{\"u}depohl}, B.~{M{\"u}ller}, H.-T. {Janka}, A.~{Marek}, G.~G. {Raffelt},
\newblock \emph{Physical Review Letters} \textbf{2010}, \emph{104}, 25 251101.

\bibitem{roberts12a}
L.~F. {Roberts}, S.~{Reddy}, G.~{Shen},
\newblock \emph{Phys. Rev. C} \textbf{2012}, \emph{86}, 6 065803.

\bibitem{martinez_pinedo12a}
G.~{Martinez-Pinedo}, T.~{Fischer}, A.~{Lohs}, L.~{Huther},
\newblock \emph{Physical Review Letters} \textbf{2012}, \emph{109}, 25 251104.

\bibitem{martinez_pinedo14}
G.~{Mart{\'\i}nez-Pinedo}, T.~{Fischer}, L.~{Huther},
\newblock \emph{Journal of Physics G Nuclear Physics} \textbf{2014}, \emph{41},
  4 044008.

\bibitem{leblanc70}
J.~M. LeBlanc, J.~R. Wilson,
\newblock \emph{ApJ} \textbf{1970}, \emph{161} 541.

\bibitem{symbalisty85}
E.~M.~D. Symbalisty, D.~N. Schramm, J.~R. Wilson,
\newblock \emph{ApJ, (Letters)} \textbf{1985}, \emph{291} L11.

\bibitem{cameron03}
A.~G.~W. {Cameron},
\newblock \emph{\apj} \textbf{2003}, \emph{587}, 1 327.

\bibitem{nishimura06}
S.~{Nishimura}, K.~{Kotake}, M.-a. {Hashimoto}, S.~{Yamada}, N.~{Nishimura},
  S.~{Fujimoto}, K.~{Sato},
\newblock \emph{\apj} \textbf{2006}, \emph{642}, 1 410.

\bibitem{moesta15}
P.~{M{\"o}sta}, C.~D. {Ott}, D.~{Radice}, L.~F. {Roberts}, E.~{Schnetter},
  R.~{Haas},
\newblock \emph{\nat} \textbf{2015}, \emph{528}, 7582 376.

\bibitem{nishimura17}
N.~{Nishimura}, H.~{Sawai}, T.~{Takiwaki}, S.~{Yamada}, F.~K. {Thielemann},
\newblock \emph{ApJ, (Letters)} \textbf{2017}, \emph{836}, 2 L21.

\bibitem{moesta18}
P.~{M{\"o}sta}, L.~F. {Roberts}, G.~{Halevi}, C.~D. {Ott}, J.~{Lippuner},
  R.~{Haas}, E.~{Schnetter},
\newblock \emph{\apj} \textbf{2018}, \emph{864}, 2 171.

\bibitem{reichert21}
M.~{Reichert}, M.~{Obergaulinger}, M.~{Eichler}, M.~{\'A}. {Aloy},
  A.~{Arcones},
\newblock \emph{\mnras} \textbf{2021}, \emph{501}, 4 5733.

\bibitem{pruet03}
J.~{Pruet}, S.~E. {Woosley}, R.~D. {Hoffman},
\newblock \emph{\apj} \textbf{2003}, \emph{586}, 2 1254.

\bibitem{pruet04}
J.~{Pruet}, T.~A. {Thompson}, R.~D. {Hoffman},
\newblock \emph{\apj} \textbf{2004}, \emph{606}, 2 1006.

\bibitem{surman06}
R.~{Surman}, G.~C. {McLaughlin}, W.~R. {Hix},
\newblock \emph{ApJ} \textbf{2006}, \emph{643} 1057.

\bibitem{siegel19a}
D.~M. {Siegel}, J.~{Barnes}, B.~D. {Metzger},
\newblock \emph{Nature} \textbf{2019}, \emph{569}, 7755 241.

\bibitem{siegel19b}
D.~M. {Siegel},
\newblock \emph{European Physical Journal A} \textbf{2019}, \emph{55}, 11 203.

\bibitem{miller20}
J.~M. {Miller}, T.~M. {Sprouse}, C.~L. {Fryer}, B.~R. {Ryan}, J.~C. {Dolence},
  M.~R. {Mumpower}, R.~{Surman},
\newblock \emph{\apj} \textbf{2020}, \emph{902}, 1 66.

\bibitem{surman04}
R.~{Surman}, G.~C. {McLaughlin},
\newblock \emph{\apj} \textbf{2004}, \emph{603}, 2 611.

\bibitem{siegel17a}
D.~M. {Siegel}, B.~D. {Metzger},
\newblock \emph{Physical Review Letters} \textbf{2017}, \emph{119}, 23 231102.

\bibitem{qian00}
Y.-Z. {Qian},
\newblock \emph{ApJL} \textbf{2000}, \emph{534} L67.

\bibitem{rosswog17a}
S.~Rosswog, U.~Feindt, O.~Korobkin, M.-R. Wu, J.~Sollerman, A.~Goobar,
  G.~Martinez-Pinedo,
\newblock \emph{Classical and Quantum Gravity} \textbf{2017}, \emph{34}, 10
  104001.

\bibitem{nakar20}
E.~{Nakar},
\newblock \emph{\physrep} \textbf{2020}, \emph{886} 1.

\bibitem{beniamini16a}
P.~{Beniamini}, K.~{Hotokezaka}, T.~{Piran},
\newblock \emph{ApJ} \textbf{2016}, \emph{832} 149.

\bibitem{ji16}
A.~P. {Ji}, A.~{Frebel}, A.~{Chiti}, J.~D. {Simon},
\newblock \emph{Nature} \textbf{2016}, \emph{531}, 7596 610.

\bibitem{hansen17a}
T.~T. {Hansen}, J.~D. {Simon}, J.~L. {Marshall}, T.~S. {Li}, D.~{Carollo},
  D.~L. {DePoy}, D.~Q. {Nagasawa}, R.~A. {Bernstein}, A.~{Drlica-Wagner}, F.~B.
  {Abdalla}, S.~{Allam}, J.~{Annis}, K.~{Bechtol}, A.~{Benoit-L{\'e}vy},
  D.~{Brooks}, E.~{Buckley-Geer}, A.~{Carnero Rosell}, M.~{Carrasco Kind},
  J.~{Carretero}, C.~E. {Cunha}, L.~N. {da Costa}, S.~{Desai}, T.~F. {Eifler},
  A.~{Fausti Neto}, B.~{Flaugher}, J.~{Frieman}, J.~{Garc{\'{\i}}a-Bellido},
  E.~{Gaztanaga}, D.~W. {Gerdes}, D.~{Gruen}, R.~A. {Gruendl}, J.~{Gschwend},
  G.~{Gutierrez}, D.~J. {James}, E.~{Krause}, K.~{Kuehn}, N.~{Kuropatkin},
  O.~{Lahav}, R.~{Miquel}, A.~A. {Plazas}, A.~K. {Romer}, E.~{Sanchez},
  B.~{Santiago}, V.~{Scarpine}, R.~C. {Smith}, M.~{Soares-Santos},
  F.~{Sobreira}, E.~{Suchyta}, M.~E.~C. {Swanson}, G.~{Tarle}, A.~R. {Walker},
  {DES Collaboration},
\newblock \emph{ApJ} \textbf{2017}, \emph{838} 44.

\bibitem{tsujimoto17}
T.~{Tsujimoto}, T.~{Matsuno}, W.~{Aoki}, M.~N. {Ishigaki}, T.~{Shigeyama},
\newblock \emph{\apjl} \textbf{2017}, \emph{850}, 1 L12.

\bibitem{wallner15a}
A.~{Wallner}, T.~{Faestermann}, J.~{Feige}, C.~{Feldstein}, K.~{Knie},
  G.~{Korschinek}, W.~{Kutschera}, A.~{Ofan}, M.~{Paul}, F.~{Quinto},
  G.~{Rugel}, P.~{Steier},
\newblock \emph{Nature Communications} \textbf{2015}, \emph{6} 5956.

\bibitem{hotokezaka15a}
K.~{Hotokezaka}, T.~{Piran}, M.~{Paul},
\newblock \emph{Nature Physics} \textbf{2015}, \emph{11} 1042.

\bibitem{macias18}
P.~{Macias}, E.~{Ramirez-Ruiz},
\newblock \emph{\apj} \textbf{2018}, \emph{860}, 2 89.

\bibitem{cowan21}
J.~J. {Cowan}, C.~{Sneden}, J.~E. {Lawler}, A.~{Aprahamian}, M.~{Wiescher},
  K.~{Langanke}, G.~{Mart{\'\i}nez-Pinedo}, F.-K. {Thielemann},
\newblock \emph{Reviews of Modern Physics} \textbf{2021}, \emph{93}, 1 015002.

\bibitem{farouqi22}
K.~{Farouqi}, F.~K. {Thielemann}, S.~{Rosswog}, K.~L. {Kratz},
\newblock \emph{A\&A} \textbf{2022}, \emph{663} A70.

\bibitem{lattimer74}
J.~M. Lattimer, D.~N. Schramm,
\newblock \emph{ApJ, (Letters)} \textbf{1974}, \emph{192} L145.

\bibitem{lattimer76}
J.~M. Lattimer, D.~N. Schramm,
\newblock \emph{ApJ} \textbf{1976}, \emph{210} 549.

\bibitem{symbalisty82}
E.~{Symbalisty}, D.~N. {Schramm},
\newblock \emph{Astrophys. Lett.} \textbf{1982}, \emph{22} 143.

\bibitem{eichler89}
D.~Eichler, M.~Livio, T.~Piran, D.~N. Schramm,
\newblock \emph{Nature} \textbf{1989}, \emph{340} 126.

\bibitem{rosswog98b}
S.~{Rosswog}, F.~K. {Thielemann}, M.~B. {Davies}, W.~{Benz}, T.~{Piran},
\newblock In W.~{Hillebrandt}, E.~{M\"uller}, editors, \emph{Nuclear
  Astrophysics}. Max-Planck-Institut f\"ur Physik und Astrophysik, Garching b.
  M\"unchen, \textbf{1998} 103.

\bibitem{rosswog99}
S.~Rosswog, M.~Liebend\"orfer, F.-K. Thielemann, M.~Davies, W.~Benz, T.~Piran,
\newblock \emph{A \&\ A} \textbf{1999}, \emph{341} 499.

\bibitem{freiburghaus99b}
C.~Freiburghaus, S.~Rosswog, F.-K. Thielemann,
\newblock \emph{ApJ} \textbf{1999}, \emph{525} L121.

\bibitem{oechslin07}
R.~{Oechslin}, H.~T. {Janka}, A.~{Marek},
\newblock \emph{\aap} \textbf{2007}, \emph{467}, 2 395.

\bibitem{bauswein13b}
A.~{Bauswein}, T.~W. {Baumgarte}, H.-T. {Janka},
\newblock \emph{Physical Review Letters} \textbf{2013}, \emph{111}, 13 131101.

\bibitem{hotokezaka13}
K.~{Hotokezaka}, K.~{Kiuchi}, K.~{Kyutoku}, H.~{Okawa}, Y.-i. {Sekiguchi},
  M.~{Shibata}, K.~{Taniguchi},
\newblock \emph{\prd} \textbf{2013}, \emph{87}, 2 024001.

\bibitem{radice18a}
D.~{Radice}, A.~{Perego}, K.~{Hotokezaka}, S.~A. {Fromm}, S.~{Bernuzzi}, L.~F.
  {Roberts},
\newblock \emph{ApJ} \textbf{2018}, \emph{869}, 2 130.

\bibitem{nedora21}
V.~{Nedora}, S.~{Bernuzzi}, D.~{Radice}, B.~{Daszuta}, A.~{Endrizzi},
  A.~{Perego}, A.~{Prakash}, M.~{Safarzadeh}, F.~{Schianchi}, D.~{Logoteta},
\newblock \emph{\apj} \textbf{2021}, \emph{906}, 2 98.

\bibitem{dessart09}
L.~{Dessart}, C.~D. {Ott}, A.~{Burrows}, S.~{Rosswog}, E.~{Livne},
\newblock \emph{ApJ} \textbf{2009}, \emph{690} 1681.

\bibitem{perego14b}
A.~{Perego}, S.~{Rosswog}, R.~M. {Cabez{\'o}n}, O.~{Korobkin},
  R.~{K{\"a}ppeli}, A.~{Arcones}, M.~{Liebend{\"o}rfer},
\newblock \emph{MNRAS} \textbf{2014}, \emph{443} 3134.

\bibitem{siegel14a}
D.~M. {Siegel}, R.~{Ciolfi}, L.~{Rezzolla},
\newblock \emph{ApJL} \textbf{2014}, \emph{785} L6.

\bibitem{fahlman18}
S.~{Fahlman}, R.~{Fern{\'a}ndez},
\newblock \emph{ApJ} \textbf{2018}, \emph{869}, 1 L3.

\bibitem{fujibayashi20}
S.~{Fujibayashi}, M.~{Shibata}, S.~{Wanajo}, K.~{Kiuchi}, K.~{Kyutoku},
  Y.~{Sekiguchi},
\newblock \emph{Phys. Rev. D} \textbf{2020}, \emph{101}, 8 083029.

\bibitem{metzger08a}
B.~D. {Metzger}, A.~L. {Piro}, E.~{Quataert},
\newblock \emph{MNRAS} \textbf{2008}, \emph{390} 781.

\bibitem{beloborodov08}
A.~M. {Beloborodov},
\newblock In {M.~Axelsson}, editor, \emph{American Institute of Physics
  Conference Series}, volume 1054 of \emph{American Institute of Physics
  Conference Series}. \textbf{2008} 51--70.

\bibitem{lee09}
W.~H. {Lee}, E.~{Ramirez-Ruiz}, D.~{L{\'o}pez-C{\'a}mara},
\newblock \emph{ApJL} \textbf{2009}, \emph{699} L93.

\bibitem{siegel18}
D.~M. {Siegel}, B.~D. {Metzger},
\newblock \emph{ApJ} \textbf{2018}, \emph{858}, 1 52.

\bibitem{fernandez19}
R.~{Fernandez}, A.~{Tchekhovskoy}, E.~{Quataert}, F.~{Foucart}, D.~{Kasen},
\newblock \emph{MNRAS} \textbf{2019}, \emph{482}, 3 3373.

\bibitem{miller19}
J.~M. {Miller}, B.~R. {Ryan}, J.~C. {Dolence}, A.~{Burrows}, C.~J. {Fontes},
  C.~L. {Fryer}, O.~{Korobkin}, J.~{Lippuner}, M.~R. {Mumpower}, R.~T.
  {Wollaeger},
\newblock \emph{\prd} \textbf{2019}, \emph{100}, 2 023008.

\bibitem{radice18b}
D.~{Radice}, A.~{Perego}, K.~{Hotokezaka}, S.~{Bernuzzi}, S.~A. {Fromm}, L.~F.
  {Roberts},
\newblock \emph{ApJL} \textbf{2018}, \emph{869}, 2 L35.

\bibitem{fujibayashi18}
S.~{Fujibayashi}, K.~{Kiuchi}, N.~{Nishimura}, Y.~{Sekiguchi}, M.~{Shibata},
\newblock \emph{ApJ} \textbf{2018}, \emph{860}, 1 64.

\bibitem{korobkin12a}
O.~{Korobkin}, S.~{Rosswog}, A.~{Arcones}, C.~{Winteler},
\newblock \emph{MNRAS} \textbf{2012}, \emph{426} 1940.

\bibitem{lippuner15}
J.~{Lippuner}, L.~F. {Roberts},
\newblock \emph{ApJ} \textbf{2015}, \emph{815}, 2 82.

\bibitem{rosswog18a}
S.~{Rosswog}, J.~{Sollerman}, U.~{Feindt}, A.~{Goobar}, O.~{Korobkin},
  R.~{Wollaeger}, C.~{Fremling}, M.~M. {Kasliwal},
\newblock \emph{A\&A} \textbf{2018}, \emph{615} A132.

\bibitem{nedora19}
V.~{Nedora}, S.~{Bernuzzi}, D.~{Radice}, A.~{Perego}, A.~{Endrizzi},
  N.~{Ortiz},
\newblock \emph{ApJ Letters} \textbf{2019}, \emph{886}, 2 L30.

\bibitem{wanajo14}
S.~{Wanajo}, Y.~{Sekiguchi}, N.~{Nishimura}, K.~{Kiuchi}, K.~{Kyutoku},
  M.~{Shibata},
\newblock \emph{ApJL} \textbf{2014}, \emph{789} L39.

\bibitem{just15}
O.~{Just}, A.~{Bauswein}, R.~A. {Pulpillo}, S.~{Goriely}, H.-T. {Janka},
\newblock \emph{MNRAS} \textbf{2015}, \emph{448} 541.

\bibitem{fernandez16b}
R.~{Fernandez}, F.~{Foucart}, D.~{Kasen}, J.~{Lippuner}, D.~{Desai}, L.~F.
  {Roberts},
\newblock \emph{ArXiv e-prints} \textbf{2016}.

\bibitem{sneden08}
C.~{Sneden}, J.~J. {Cowan}, R.~{Gallino},
\newblock \emph{Annual Review of Astronomy and Astrophysics} \textbf{2008},
  \emph{46} 241.

\bibitem{siegel22}
D.~M. {Siegel},
\newblock \emph{Nature Reviews Physics} \textbf{2022}, \emph{4}, 5 306.

\bibitem{metzger19a}
B.~D. {Metzger},
\newblock \emph{Living Reviews in Relativity} \textbf{2019}, \emph{23}, 1 1.

\bibitem{margutti21}
R.~{Margutti}, R.~{Chornock},
\newblock \emph{Annual Reviews of Astronomy and Astrophysics} \textbf{2021},
  \emph{59} 155.

\bibitem{pian21}
E.~{Pian},
\newblock \emph{Frontiers in Astronomy and Space Sciences} \textbf{2021},
  \emph{7} 108.

\bibitem{shibata19a}
M.~{Shibata}, K.~{Hotokezaka},
\newblock \emph{Annual Review of Nuclear and Particle Science} \textbf{2019},
  \emph{69}, 1 annurev.

\bibitem{radice20}
D.~{Radice}, S.~{Bernuzzi}, A.~{Perego},
\newblock \emph{Annual Review of Nuclear and Particle Science} \textbf{2020},
  \emph{70} 95.

\bibitem{abbott17b}
B.~P. {Abbott}, R.~{Abbott}, T.~D. {Abbott}, F.~{Acernese}, K.~{Ackley},
  C.~{Adams}, T.~{Adams}, P.~{Addesso}, R.~X. {Adhikari}, V.~B. {Adya}, et~al.,
\newblock \emph{Physical Review Letters} \textbf{2017}, \emph{119}, 16 161101.

\bibitem{abbott17c}
B.~P. {Abbott}, R.~{Abbott}, T.~D. {Abbott}, F.~{Acernese}, K.~{Ackley},
  C.~{Adams}, T.~{Adams}, P.~{Addesso}, R.~X. {Adhikari}, V.~B. {Adya}, et~al.,
\newblock \emph{ApJL} \textbf{2017}, \emph{848} L12.

\bibitem{hotokezaka11}
K.~{Hotokezaka}, K.~{Kyutoku}, H.~{Okawa}, M.~{Shibata}, K.~{Kiuchi},
\newblock \emph{Phys. Rev. D} \textbf{2011}, \emph{83}, 12 124008.

\bibitem{koeppel19}
S.~{K{\"o}ppel}, L.~{Bovard}, L.~{Rezzolla},
\newblock \emph{ApJL} \textbf{2019}, \emph{872}, 1 L16.

\bibitem{bernuzzi20b}
S.~{Bernuzzi}, M.~{Breschi}, B.~{Daszuta}, A.~{Endrizzi}, D.~{Logoteta},
  V.~{Nedora}, A.~{Perego}, D.~{Radice}, F.~{Schianchi}, F.~{Zappa},
  I.~{Bombaci}, N.~{Ortiz},
\newblock \emph{\mnras} \textbf{2020}, \emph{497}, 2 1488.

\bibitem{kashyap22}
R.~{Kashyap}, A.~{Das}, D.~{Radice}, S.~{Padamata}, A.~{Prakash},
  D.~{Logoteta}, A.~{Perego}, D.~A. {Godzieba}, S.~{Bernuzzi}, I.~{Bombaci},
  F.~J. {Fattoyev}, B.~T. {Reed}, A.~{da Silva Schneider},
\newblock \emph{arXiv e-prints} \textbf{2021}, arXiv:2111.05183.

\bibitem{shibata06c}
M.~{Shibata}, K.~{Taniguchi},
\newblock \emph{Phys. Rev. D} \textbf{2006}, \emph{73}, 6 064027.

\bibitem{kiuchi09}
K.~{Kiuchi}, Y.~{Sekiguchi}, M.~{Shibata}, K.~{Taniguchi},
\newblock \emph{\prd} \textbf{2009}, \emph{80}, 6 064037.

\bibitem{hotokezaka13b}
K.~{Hotokezaka}, K.~{Kiuchi}, K.~{Kyutoku}, T.~{Muranushi}, Y.-i. {Sekiguchi},
  M.~{Shibata}, K.~{Taniguchi},
\newblock \emph{Phys. Rev. D} \textbf{2013}, \emph{88}, 4 044026.

\bibitem{tauris17}
T.~M. {Tauris}, M.~{Kramer}, P.~C.~C. {Freire}, N.~{Wex}, H.-T. {Janka},
  N.~{Langer}, P.~{Podsiadlowski}, E.~{Bozzo}, S.~{Chaty}, M.~U. {Kruckow},
  E.~P.~J. {van den Heuvel}, J.~{Antoniadis}, R.~P. {Breton}, D.~J. {Champion},
\newblock \emph{ApJ} \textbf{2017}, \emph{846} 170.

\bibitem{hotokezaka13a}
K.~{Hotokezaka}, K.~{Kiuchi}, K.~{Kyutoku}, H.~{Okawa}, Y.-i. {Sekiguchi},
  M.~{Shibata}, K.~{Taniguchi},
\newblock \emph{Phys. Rev. D} \textbf{2013}, \emph{87}, 2 024001.

\bibitem{bauswein13a}
A.~{Bauswein}, S.~{Goriely}, H.-T. {Janka},
\newblock \emph{ApJ} \textbf{2013}, \emph{773} 78.

\bibitem{rosswog22b}
S.~{Rosswog}, P.~{Diener}, F.~{Torsello},
\newblock \emph{Symmetry} \textbf{2022}, \emph{14}, 6 1280.

\bibitem{dietrich17}
T.~{Dietrich}, M.~{Ujevic},
\newblock \emph{Classical and Quantum Gravity} \textbf{2017}, \emph{34}, 10
  105014.

\bibitem{price06}
D.~Price, S.~Rosswog,
\newblock \emph{Science} \textbf{2006}, \emph{312} 719.

\bibitem{kiuchi15}
K.~{Kiuchi}, P.~{Cerd{\'a}-Dur{\'a}n}, K.~{Kyutoku}, Y.~{Sekiguchi},
  M.~{Shibata},
\newblock \emph{Phys. Rev. D} \textbf{2015}, \emph{92}, 12 124034.

\bibitem{palenzuela22}
C.~{Palenzuela}, R.~{Aguilera-Miret}, F.~{Carrasco}, R.~{Ciolfi}, J.~V.
  {Kalinani}, W.~{Kastaun}, B.~{Mi{\~n}ano}, D.~{Vigan{\`o}},
\newblock \emph{\prd} \textbf{2022}, \emph{106}, 2 023013.

\bibitem{wollaeger19}
R.~T. {Wollaeger}, C.~L. {Fryer}, C.~J. {Fontes}, J.~{Lippuner}, W.~T.
  {Vestrand}, M.~R. {Mumpower}, O.~{Korobkin}, A.~L. {Hungerford}, W.~P.
  {Even},
\newblock \emph{\apj} \textbf{2019}, \emph{880}, 1 22.

\bibitem{sekiguchi16a}
Y.~{Sekiguchi}, K.~{Kiuchi}, K.~{Kyutoku}, M.~{Shibata}, K.~{Taniguchi},
\newblock \emph{ArXiv e-prints} \textbf{2016}.

\bibitem{rosswog13b}
S.~{Rosswog},
\newblock \emph{Royal Society of London Philosophical Transactions Series A}
  \textbf{2013}, \emph{371} 20272.

\bibitem{lehner16a}
L.~{Lehner}, S.~L. {Liebling}, C.~{Palenzuela}, O.~L. {Caballero},
  E.~{O'Connor}, M.~{Anderson}, D.~{Neilsen},
\newblock \emph{Classical and Quantum Gravity} \textbf{2016}, \emph{33}, 18
  184002.

\bibitem{kyutoku14}
K.~{Kyutoku}, K.~{Ioka}, M.~{Shibata},
\newblock \emph{\mnras} \textbf{2014}, \emph{437}, 1 L6.

\bibitem{metzger15a}
B.~D. {Metzger}, A.~{Bauswein}, S.~{Goriely}, D.~{Kasen},
\newblock \emph{MNRAS} \textbf{2015}, \emph{446} 1115.

\bibitem{dean21}
C.~{Dean}, R.~{Fern{\'a}ndez}, B.~D. {Metzger},
\newblock \emph{ApJ} \textbf{2021}, \emph{921}, 2 161.

\bibitem{combi22}
L.~{Combi}, D.~{Siegel},
\newblock \emph{arXiv e-prints} \textbf{2022}, arXiv:2206.03618.

\bibitem{kulkarni05}
S.~R. {Kulkarni},
\newblock \emph{ArXiv Astrophysics e-prints} \textbf{2005}.

\bibitem{beloborodov20}
A.~M. {Beloborodov}, C.~{Lundman}, Y.~{Levin},
\newblock \emph{ApJ} \textbf{2020}, \emph{897}, 2 141.

\bibitem{nakar11a}
E.~{Nakar}, T.~{Piran},
\newblock \emph{Nature} \textbf{2011}, \emph{478} 82.

\bibitem{mooley17}
K.~P. {Mooley}, E.~{Nakar}, K.~{Hotokezaka}, G.~{Hallinan}, A.~{Corsi}, D.~A.
  {Frail}, A.~{Horesh}, T.~{Murphy}, E.~{Lenc}, D.~L. {Kaplan}, K.~{de},
  D.~{Dobie}, P.~{Chandra}, A.~{Deller}, O.~{Gottlieb}, M.~M. {Kasliwal}, S.~R.
  {Kulkarni}, S.~T. {Myers}, S.~{Nissanke}, T.~{Piran}, C.~{Lynch},
  V.~{Bhalerao}, S.~{Bourke}, K.~W. {Bannister}, L.~P. {Singer},
\newblock \emph{Nature} \textbf{2018}, \emph{554} 207.

\bibitem{hotokezaka18}
K.~{Hotokezaka}, K.~{Kiuchi}, M.~{Shibata}, E.~{Nakar}, T.~{Piran},
\newblock \emph{\apj} \textbf{2018}, \emph{867}, 2 95.

\bibitem{hajela22}
A.~{Hajela}, R.~{Margutti}, J.~S. {Bright}, K.~D. {Alexander}, B.~D. {Metzger},
  V.~{Nedora}, A.~{Kathirgamaraju}, B.~{Margalit}, D.~{Radice}, C.~{Guidorzi},
  E.~{Berger}, A.~{MacFadyen}, D.~{Giannios}, R.~{Chornock}, I.~{Heywood},
  L.~{Sironi}, O.~{Gottlieb}, D.~{Coppejans}, T.~{Laskar}, Y.~{Cendes}, R.~B.
  {Duran}, T.~{Eftekhari}, W.~{Fong}, A.~{McDowell}, M.~{Nicholl}, X.~{Xie},
  J.~{Zrake}, S.~{Bernuzzi}, F.~S. {Broekgaarden}, C.~D. {Kilpatrick},
  G.~{Terreran}, V.~A. {Villar}, P.~K. {Blanchard}, S.~{Gomez},
  G.~{Hosseinzadeh}, D.~J. {Matthews}, J.~C. {Rastinejad},
\newblock \emph{ApJL} \textbf{2022}, \emph{927}, 1 L17.

\bibitem{bildsten92}
L.~{Bildsten}, C.~{Cutler},
\newblock \emph{ApJ} \textbf{1992}, \emph{400} 175.

\bibitem{kochanek92}
C.~S. {Kochanek},
\newblock \emph{ApJ} \textbf{1992}, \emph{398} 234.

\bibitem{papenfort21}
L.~J. {Papenfort}, E.~R. {Most}, S.~{Tootle}, L.~{Rezzolla},
\newblock \emph{\mnras} \textbf{2022}, \emph{513}, 3 3646.

\bibitem{rosswog00}
S.~{Rosswog}, M.~B. {Davies}, F.-K. {Thielemann}, T.~{Piran},
\newblock \emph{A\&A} \textbf{2000}, \emph{360} 171.

\bibitem{chaurasia20}
S.~V. {Chaurasia}, T.~{Dietrich}, M.~{Ujevic}, K.~{Hendriks}, R.~{Dudi}, F.~M.
  {Fabbri}, W.~{Tichy}, B.~{Br{\"u}gmann},
\newblock \emph{\prd} \textbf{2020}, \emph{102}, 2 024087.

\bibitem{rosswog21a}
S.~{Rosswog}, P.~{Diener},
\newblock \emph{Classical and Quantum Gravity} \textbf{2021}, \emph{38}, 11
  115002.

\bibitem{diener22a}
P.~{Diener}, S.~{Rosswog}, F.~{Torsello},
\newblock \emph{European Journal of Physics A, Volume 58, Issue 4, article
  id.74} \textbf{2022}.

\bibitem{rosswog22a}
S.~{Rosswog},
\newblock \emph{Cambridge University Press} \textbf{2022}, arXiv:2201.05896.

\bibitem{qian96b}
Y.-Z. {Qian}, S.~E. {Woosley},
\newblock \emph{ApJ} \textbf{1996}, \emph{471} 331.

\bibitem{martinez_pinedo12}
G.~{Mart{\'\i}nez-Pinedo}, T.~{Fischer}, A.~{Lohs}, L.~{Huther},
\newblock \emph{Phys. Rev. Lett.} \textbf{2012}, \emph{109}, 25 251104.

\bibitem{martin15}
D.~{Martin}, A.~{Perego}, A.~{Arcones}, F.-K. {Thielemann}, O.~{Korobkin},
  S.~{Rosswog},
\newblock \emph{ApJ} \textbf{2015}, \emph{813} 2.

\bibitem{ciolfi20}
R.~{Ciolfi}, J.~V. {Kalinani},
\newblock \emph{ApJ Letters} \textbf{2020}, \emph{900}, 2 L35.

\bibitem{evans17}
P.~A. {Evans}, S.~B. {Cenko}, J.~A. {Kennea}, S.~W.~K. {Emery}, N.~P.~M.
  {Kuin}, O.~{Korobkin}, R.~T. {Wollaeger}, C.~L. {Fryer}, K.~K. {Madsen},
  F.~A. {Harrison}, Y.~{Xu}, E.~{Nakar}, K.~{Hotokezaka}, A.~{Lien},
  S.~{Campana}, S.~R. {Oates}, E.~{Troja}, A.~A. {Breeveld}, F.~E. {Marshall},
  S.~D. {Barthelmy}, A.~P. {Beardmore}, D.~N. {Burrows}, G.~{Cusumano},
  A.~{D'Ai}, P.~{D'Avanzo}, V.~{D'Elia}, M.~{de Pasquale}, W.~P. {Even}, C.~J.
  {Fontes}, K.~{Forster}, J.~{Garcia}, P.~{Giommi}, B.~{Grefenstette},
  C.~{Gronwall}, D.~H. {Hartmann}, M.~{Heida}, A.~L. {Hungerford}, M.~M.
  {Kasliwal}, H.~A. {Krimm}, A.~J. {Levan}, D.~{Malesani}, A.~{Melandri},
  H.~{Miyasaka}, J.~A. {Nousek}, P.~T. {O'Brien}, J.~P. {Osborne}, C.~{Pagani},
  K.~L. {Page}, D.~M. {Palmer}, M.~{Perri}, S.~{Pike}, J.~L. {Racusin},
  S.~{Rosswog}, M.~H. {Siegel}, T.~{Sakamoto}, B.~{Sbarufatti},
  G.~{Tagliaferri}, N.~R. {Tanvir}, A.~{Tohuvavohu},
\newblock \emph{Science} \textbf{2017}, \emph{358} 1565.

\bibitem{nicholl17}
M.~{Nicholl}, E.~{Berger}, D.~{Kasen}, B.~D. {Metzger}, J.~{Elias},
  C.~{Brice{\~n}o}, K.~D. {Alexander}, P.~K. {Blanchard}, R.~{Chornock}, P.~S.
  {Cowperthwaite}, T.~{Eftekhari}, W.~{Fong}, R.~{Margutti}, V.~A. {Villar},
  P.~K.~G. {Williams}, W.~{Brown}, J.~{Annis}, A.~{Bahramian}, D.~{Brout},
  D.~A. {Brown}, H.-Y. {Chen}, J.~C. {Clemens}, E.~{Dennihy}, B.~{Dunlap},
  D.~E. {Holz}, E.~{Marchesini}, F.~{Massaro}, N.~{Moskowitz}, I.~{Pelisoli},
  A.~{Rest}, F.~{Ricci}, M.~{Sako}, M.~{Soares-Santos}, J.~{Strader},
\newblock \emph{ApJL} \textbf{2017}, \emph{848} L18.

\bibitem{mccully17}
C.~{McCully}, D.~{Hiramatsu}, D.~A. {Howell}, G.~{Hosseinzadeh}, I.~{Arcavi},
  D.~{Kasen}, J.~{Barnes}, M.~M. {Shara}, T.~B. {Williams},
  P.~{V{\"a}is{\"a}nen}, S.~B. {Potter}, E.~{Romero-Colmenero}, S.~M.
  {Crawford}, D.~A.~H. {Buckley}, J.~{Cooke}, I.~{Andreoni}, T.~A. {Pritchard},
  J.~{Mao}, M.~{Gromadzki}, J.~{Burke},
\newblock \emph{\apjl} \textbf{2017}, \emph{848} L32.

\bibitem{metzger18a}
B.~D. {Metzger}, T.~A. {Thompson}, E.~{Quataert},
\newblock \emph{\apj} \textbf{2018}, \emph{856} 101.

\bibitem{korobkin21}
O.~{Korobkin}, R.~T. {Wollaeger}, C.~L. {Fryer}, A.~L. {Hungerford},
  S.~{Rosswog}, C.~J. {Fontes}, M.~R. {Mumpower}, E.~A. {Chase}, W.~P. {Even},
  J.~{Miller}, G.~W. {Misch}, J.~{Lippuner},
\newblock \emph{ApJ} \textbf{2021}, \emph{910}, 2 116.

\bibitem{cowperthwaite17}
P.~S. {Cowperthwaite}, E.~{Berger}, V.~A. {Villar}, B.~D. {Metzger et al.},
\newblock \emph{ApJL} \textbf{2017}, \emph{848} L17.

\bibitem{beloborodov03}
A.~M. {Beloborodov},
\newblock \emph{ApJ} \textbf{2003}, \emph{588}, 2 931.

\bibitem{kawanaka07}
N.~{Kawanaka}, S.~{Mineshige},
\newblock \emph{ApJ} \textbf{2007}, \emph{662}, 2 1156.

\bibitem{chen07}
W.~{Chen}, A.~M. {Beloborodov},
\newblock \emph{ApJ} \textbf{2007}, \emph{657} 383.

\bibitem{metzger09b}
B.~D. {Metzger}, A.~L. {Piro}, E.~{Quataert},
\newblock \emph{MNRAS} \textbf{2009}, \emph{396} 304.

\bibitem{roederer10}
I.~U. {Roederer}, J.~J. {Cowan}, A.~I. {Karakas}, K.-L. {Kratz}, M.~{Lugaro},
  J.~{Simmerer}, K.~{Farouqi}, C.~{Sneden},
\newblock \emph{\apj} \textbf{2010}, \emph{724}, 2 975.

\bibitem{holmbeck18}
E.~M. {Holmbeck}, T.~C. {Beers}, I.~U. {Roederer}, V.~M. {Placco}, T.~T.
  {Hansen}, C.~M. {Sakari}, C.~{Sneden}, C.~{Liu}, Y.~S. {Lee}, J.~J. {Cowan},
  A.~{Frebel},
\newblock \emph{ApJ} \textbf{2018}, \emph{859} L24.

\bibitem{holmbeck19a}
E.~M. {Holmbeck}, T.~M. {Sprouse}, M.~R. {Mumpower}, N.~{Vassh}, R.~{Surman},
  T.~C. {Beers}, T.~{Kawano},
\newblock \emph{ApJ} \textbf{2019}, \emph{870} 23.

\bibitem{wu17}
M.-R. {Wu}, I.~{Tamborra}, O.~{Just}, H.-T. {Janka},
\newblock \emph{\prd} \textbf{2017}, \emph{96}, 12 123015.

\bibitem{eichler19}
M.~{Eichler}, W.~{Sayar}, A.~{Arcones}, T.~{Rauscher},
\newblock \emph{\apj} \textbf{2019}, \emph{879}, 1 47.

\bibitem{holmbeck19b}
E.~M. {Holmbeck}, A.~{Frebel}, G.~C. {McLaughlin}, M.~R. {Mumpower}, T.~M.
  {Sprouse}, R.~{Surman},
\newblock \emph{ApJ} \textbf{2019}, \emph{881} 5.

\bibitem{miller19b}
J.~M. {Miller}, B.~R. {Ryan}, J.~C. {Dolence},
\newblock \emph{ApJS} \textbf{2019}, \emph{241}, 2 30.

\bibitem{just22}
O.~{Just}, S.~{Goriely}, H.~T. {Janka}, S.~{Nagataki}, A.~{Bauswein},
\newblock \emph{MNRAS} \textbf{2022}, \emph{509}, 1 1377.

\bibitem{artemova96}
I.~V. {Artemova}, G.~{Bjoernsson}, I.~D. {Novikov},
\newblock \emph{ApJ} \textbf{1996}, \emph{461} 565.

\bibitem{fernandez20}
R.~{Fern{\'a}ndez}, F.~{Foucart}, J.~{Lippuner},
\newblock \emph{\mnras} \textbf{2020}, \emph{497}, 3 3221.

\bibitem{abbott17h}
B.~P. {Abbott}, R.~{Abbott}, T.~D. {Abbott}, F.~{Acernese}, K.~{Ackley},
  C.~{Adams}, T.~{Adams}, P.~{Addesso}, R.~X. {Adhikari}, V.~B. {Adya}, et~al.,
\newblock \emph{Physical Review Letters} \textbf{2017}, \emph{119}, 16 161101.

\bibitem{li98}
L.-X. {Li}, B.~{Paczy{\'n}ski},
\newblock \emph{ApJL} \textbf{1998}, \emph{507} L59.

\bibitem{metzger10b}
B.~D. {Metzger}, G.~{Martinez-Pinedo}, S.~{Darbha}, E.~{Quataert},
  A.~{Arcones}, D.~{Kasen}, R.~{Thomas}, P.~{Nugent}, I.~V. {Panov}, N.~T.
  {Zinner},
\newblock \emph{MNRAS} \textbf{2010}, \emph{406} 2650.

\bibitem{kasen13a}
D.~{Kasen}, N.~R. {Badnell}, J.~{Barnes},
\newblock \emph{ApJ} \textbf{2013}, \emph{774} 25.

\bibitem{tanaka13a}
M.~{Tanaka}, K.~{Hotokezaka},
\newblock \emph{ApJ} \textbf{2013}, \emph{775} 113.

\bibitem{fontes15}
C.~J. {Fontes}, C.~L. {Fryer}, A.~L. {Hungerford}, P.~{Hakel}, J.~{Colgan},
  D.~P. {Kilcrease}, M.~E. {Sherrill},
\newblock \emph{High Energy Density Physics} \textbf{2015}, \emph{16} 53.

\bibitem{kasen17}
D.~{Kasen}, B.~{Metzger}, J.~{Barnes}, E.~{Quataert}, E.~{Ramirez-Ruiz},
\newblock \emph{Nature} \textbf{2017}, \emph{551} 80.

\bibitem{fontes17a}
C.~J. {Fontes}, C.~L. {Fryer}, A.~L. {Hungerford}, R.~T. {Wollaeger},
  S.~{Rosswog}, E.~{Berger},
\newblock \emph{ArXiv e-prints} \textbf{2017}.

\bibitem{even20}
W.~{Even}, O.~{Korobkin}, C.~L. {Fryer}, C.~J. {Fontes}, R.~T. {Wollaeger},
  A.~{Hungerford}, J.~{Lippuner}, J.~{Miller}, M.~R. {Mumpower}, G.~W. {Misch},
\newblock \emph{\apj} \textbf{2020}, \emph{899}, 1 24.

\bibitem{wollaeger18}
R.~T. {Wollaeger}, O.~{Korobkin}, C.~J. {Fontes}, S.~K. {Rosswog}, W.~P.
  {Even}, C.~L. {Fryer}, J.~{Sollerman}, A.~L. {Hungerford}, D.~R. {van
  Rossum}, A.~B. {Wollaber},
\newblock \emph{\mnras} \textbf{2018}, \emph{478} 3298.

\bibitem{fontes20}
C.~J. {Fontes}, C.~L. {Fryer}, A.~L. {Hungerford}, R.~T. {Wollaeger},
  O.~{Korobkin},
\newblock \emph{MNRAS} \textbf{2020}, \emph{493}, 3 4143.

\bibitem{hotokezaka21}
K.~{Hotokezaka}, M.~{Tanaka}, D.~{Kato}, G.~{Gaigalas},
\newblock \emph{\mnras} \textbf{2021}, \emph{506}, 4 5863.

\bibitem{pognan22a}
Q.~{Pognan}, A.~{Jerkstrand}, J.~{Grumer},
\newblock \emph{\mnras} \textbf{2022}, \emph{513}, 4 5174.

\bibitem{pognan22b}
Q.~{Pognan}, A.~{Jerkstrand}, J.~{Grumer},
\newblock \emph{\mnras} \textbf{2022}, \emph{510}, 3 3806.

\bibitem{chornock17}
R.~{Chornock}, E.~{Berger}, D.~{Kasen}, P.~S. {Cowperthwaite}, M.~{Nicholl},
  V.~A. {Villar}, K.~D. {Alexander}, P.~K. {Blanchard}, T.~{Eftekhari},
  W.~{Fong}, R.~{Margutti}, P.~K.~G. {Williams}, J.~{Annis}, D.~{Brout}, D.~A.
  {Brown}, H.-Y. {Chen}, M.~R. {Drout}, B.~{Farr}, R.~J. {Foley}, J.~A.
  {Frieman}, C.~L. {Fryer}, K.~{Herner}, D.~E. {Holz}, R.~{Kessler},
  T.~{Matheson}, B.~D. {Metzger}, E.~{Quataert}, A.~{Rest}, M.~{Sako}, D.~M.
  {Scolnic}, N.~{Smith}, M.~{Soares-Santos},
\newblock \emph{\apjl} \textbf{2017}, \emph{848} L19.

\bibitem{pian17}
E.~{Pian}, P.~{D'Avanzo}, S.~{Benetti}, M.~{Branchesi}, E.~{Brocato},
  S.~{Campana}, E.~{Cappellaro}, S.~{Covino}, V.~{D'Elia}, J.~P.~U. {Fynbo},
  F.~{Getman}, G.~{Ghirlanda}, G.~{Ghisellini}, A.~{Grado}, G.~{Greco},
  J.~{Hjorth}, C.~{Kouveliotou}, A.~{Levan}, L.~{Limatola}, D.~{Malesani},
  P.~A. {Mazzali}, A.~{Melandri}, P.~{M{\o}ller}, L.~{Nicastro}, E.~{Palazzi},
  S.~{Piranomonte}, A.~{Rossi}, O.~S. {Salafia}, J.~{Selsing}, G.~{Stratta},
  M.~{Tanaka}, N.~R. {Tanvir}, L.~{Tomasella}, D.~{Watson}, S.~{Yang},
  L.~{Amati}, L.~A. {Antonelli}, S.~{Ascenzi}, M.~G. {Bernardini},
  M.~{Bo{\"e}r}, F.~{Bufano}, A.~{Bulgarelli}, M.~{Capaccioli}, P.~{Casella},
  A.~J. {Castro-Tirado}, E.~{Chassande-Mottin}, R.~{Ciolfi}, C.~M.
  {Copperwheat}, M.~{Dadina}, G.~{De Cesare}, A.~{di Paola}, Y.~Z. {Fan},
  B.~{Gendre}, G.~{Giuffrida}, A.~{Giunta}, L.~K. {Hunt}, G.~L. {Israel}, Z.-P.
  {Jin}, M.~M. {Kasliwal}, S.~{Klose}, M.~{Lisi}, F.~{Longo}, E.~{Maiorano},
  M.~{Mapelli}, N.~{Masetti}, L.~{Nava}, B.~{Patricelli}, D.~{Perley},
  A.~{Pescalli}, T.~{Piran}, A.~{Possenti}, L.~{Pulone}, M.~{Razzano},
  R.~{Salvaterra}, P.~{Schipani}, M.~{Spera}, A.~{Stamerra}, L.~{Stella},
  G.~{Tagliaferri}, V.~{Testa}, E.~{Troja}, M.~{Turatto}, S.~D. {Vergani},
  D.~{Vergani},
\newblock \emph{\nat} \textbf{2017}, \emph{551} 67.

\bibitem{tanvir17}
N.~R. {Tanvir}, A.~J. {Levan}, C.~{Gonz{\'a}lez-Fern{\'a}ndez}, O.~{Korobkin},
  I.~{Mandel}, S.~{Rosswog}, J.~{Hjorth}, P.~{D'Avanzo}, A.~S. {Fruchter},
  C.~L. {Fryer}, T.~{Kangas}, B.~{Milvang-Jensen}, S.~{Rosetti}, D.~{Steeghs},
  R.~T. {Wollaeger}, Z.~{Cano}, C.~M. {Copperwheat}, S.~{Covino}, V.~{D'Elia},
  A.~{de Ugarte Postigo}, P.~A. {Evans}, W.~P. {Even}, S.~{Fairhurst},
  R.~{Figuera Jaimes}, C.~J. {Fontes}, Y.~I. {Fujii}, J.~P.~U. {Fynbo}, B.~P.
  {Gompertz}, J.~{Greiner}, G.~{Hodosan}, M.~J. {Irwin}, P.~{Jakobsson}, U.~G.
  {J{\o}rgensen}, D.~A. {Kann}, J.~D. {Lyman}, D.~{Malesani}, R.~G. {McMahon},
  A.~{Melandri}, P.~T. {O'Brien}, J.~P. {Osborne}, E.~{Palazzi}, D.~A.
  {Perley}, E.~{Pian}, S.~{Piranomonte}, M.~{Rabus}, E.~{Rol}, A.~{Rowlinson},
  S.~{Schulze}, P.~{Sutton}, C.~C. {Th{\"o}ne}, K.~{Ulaczyk}, D.~{Watson},
  K.~{Wiersema}, R.~A.~M.~J. {Wijers},
\newblock \emph{ApJL} \textbf{2017}, \emph{848} L27.

\bibitem{kawaguchi18}
K.~{Kawaguchi}, M.~{Shibata}, M.~{Tanaka},
\newblock \emph{\apj} \textbf{2018}, \emph{865} L21.

\bibitem{piro18}
A.~L. {Piro}, J.~A. {Kollmeier},
\newblock \emph{\apj} \textbf{2018}, \emph{855}, 2 103.

\bibitem{kisaka16}
S.~{Kisaka}, K.~{Ioka}, E.~{Nakar},
\newblock \emph{\apj} \textbf{2016}, \emph{818}, 2 104.

\bibitem{troja17}
E.~{Troja}, L.~{Piro}, H.~{van Eerten}, R.~T. {Wollaeger}, M.~{Im}, O.~D.
  {Fox}, N.~R. {Butler}, S.~B. {Cenko}, T.~{Sakamoto}, C.~L. {Fryer},
  R.~{Ricci}, A.~{Lien}, R.~E. {Ryan}, O.~{Korobkin}, S.~K. {Lee}, many more,
\newblock \emph{Nature} \textbf{2017}, \emph{551}, 7678 71.

\bibitem{kasliwal17}
M.~M. {Kasliwal}, E.~{Nakar}, L.~P. {Singer}, D.~L. {Kaplan}, et~al.,
\newblock \emph{Science} \textbf{2017}, \emph{358} 1559.

\bibitem{watson19}
D.~{Watson}, C.~J. {Hansen}, J.~{Selsing}, A.~{Koch}, D.~B. {Malesani}, A.~C.
  {Andersen}, J.~P.~U. {Fynbo}, A.~{Arcones}, A.~{Bauswein}, S.~{Covino},
  A.~{Grado}, K.~E. {Heintz}, L.~{Hunt}, C.~{Kouveliotou}, G.~{Leloudas}, A.~J.
  {Levan}, P.~{Mazzali}, E.~{Pian},
\newblock \emph{Nature} \textbf{2019}, \emph{574}, 7779 497.

\bibitem{domoto21}
N.~{Domoto}, M.~{Tanaka}, S.~{Wanajo}, K.~{Kawaguchi},
\newblock \emph{\apj} \textbf{2021}, \emph{913}, 1 26.

\bibitem{gillanders21}
J.~H. {Gillanders}, M.~{McCann}, S.~A. {Sim}, S.~J. {Smartt}, C.~P. {Ballance},
\newblock \emph{\mnras} \textbf{2021}, \emph{506}, 3 3560.

\bibitem{gillanders22}
J.~H. {Gillanders}, S.~J. {Smartt}, S.~A. {Sim}, A.~{Bauswein}, S.~{Goriely},
\newblock \emph{\mnras} \textbf{2022}.

\bibitem{perego22}
A.~{Perego}, D.~{Vescovi}, A.~{Fiore}, L.~{Chiesa}, C.~{Vogl}, S.~{Benetti},
  S.~{Bernuzzi}, M.~{Branchesi}, E.~{Cappellaro}, S.~{Cristallo},
  A.~{Fl{\"o}rs}, W.~E. {Kerzendorf}, D.~{Radice},
\newblock \emph{\apj} \textbf{2022}, \emph{925}, 1 22.

\bibitem{drout17}
M.~R. {Drout}, A.~L. {Piro}, B.~J. {Shappee}, C.~D. {Kilpatrick}, J.~D.
  {Simon}, C.~{Contreras}, D.~A. {Coulter}, R.~J. {Foley}, M.~R. {Siebert},
  N.~{Morrell}, K.~{Boutsia}, F.~{Di Mille}, T.~W.-S. {Holoien}, D.~{Kasen},
  J.~A. {Kollmeier}, B.~F. {Madore}, A.~J. {Monson}, A.~{Murguia-Berthier},
  Y.-C. {Pan}, J.~X. {Prochaska}, E.~{Ramirez-Ruiz}, A.~{Rest}, C.~{Adams},
  K.~{Alatalo}, E.~{Ba{\~n}ados}, J.~{Baughman}, T.~C. {Beers}, R.~A.
  {Bernstein}, T.~{Bitsakis}, A.~{Campillay}, T.~T. {Hansen}, C.~R. {Higgs},
  A.~P. {Ji}, G.~{Maravelias}, J.~L. {Marshall}, C.~{Moni Bidin}, J.~L.
  {Prieto}, K.~C. {Rasmussen}, C.~{Rojas-Bravo}, A.~L. {Strom}, N.~{Ulloa},
  J.~{Vargas-Gonz{\'a}lez}, Z.~{Wan}, D.~D. {Whitten},
\newblock \emph{Science, in press, available via doi:10.1126/science.aaq0049}
  \textbf{2017}.

\bibitem{wu19}
M.-R. {Wu}, J.~{Barnes}, G.~{Mart{\'\i}nez-Pinedo}, B.~D. {Metzger},
\newblock \emph{Phys. Rev. Lett.} \textbf{2019}, \emph{122}, 6 062701.

\bibitem{kasliwal22}
M.~M. {Kasliwal}, D.~{Kasen}, R.~M. {Lau}, D.~A. {Perley}, S.~{Rosswog}, E.~O.
  {Ofek}, K.~{Hotokezaka}, R.-R. {Chary}, J.~{Sollerman}, A.~{Goobar}, D.~L.
  {Kaplan},
\newblock \emph{\mnras} \textbf{2022}, \emph{510}, 1 L7.

\bibitem{villar18}
V.~A. {Villar}, P.~S. {Cowperthwaite}, E.~{Berger}, P.~K. {Blanchard},
  S.~{Gomez}, K.~D. {Alexander}, R.~{Margutti}, R.~{Chornock}, T.~{Eftekhari},
  G.~G. {Fazio}, J.~{Guillochon}, J.~L. {Hora}, M.~{Nicholl}, P.~K.~G.
  {Williams},
\newblock \emph{\apjl} \textbf{2018}, \emph{862}, 1 L11.

\bibitem{domoto22}
N.~{Domoto}, M.~{Tanaka}, D.~{Kato}, K.~{Kawaguchi}, K.~{Hotokezaka},
  S.~{Wanajo},
\newblock \emph{arXiv e-prints} \textbf{2022}, arXiv:2206.04232.

\bibitem{shappee17}
B.~J. {Shappee}, J.~D. {Simon}, M.~R. {Drout}, A.~L. {Piro}, N.~{Morrell},
  J.~L. {Prieto}, D.~{Kasen}, T.~W.-S. {Holoien}, J.~A. {Kollmeier}, D.~D.
  {Kelson}, D.~A. {Coulter}, R.~J. {Foley}, C.~D. {Kilpatrick}, M.~R.
  {Siebert}, B.~F. {Madore}, A.~{Murguia-Berthier}, Y.-C. {Pan}, J.~X.
  {Prochaska}, E.~{Ramirez-Ruiz}, A.~{Rest}, C.~{Adams}, K.~{Alatalo},
  E.~{Ba{\~n}ados}, J.~{Baughman}, R.~A. {Bernstein}, T.~{Bitsakis},
  K.~{Boutsia}, J.~R. {Bravo}, F.~{Di Mille}, C.~R. {Higgs}, A.~P. {Ji},
  G.~{Maravelias}, J.~L. {Marshall}, V.~M. {Placco}, G.~{Prieto}, Z.~{Wan},
\newblock \emph{Science} \textbf{2017}, \emph{358} 1574.

\bibitem{villar17}
V.~A. {Villar}, J.~{Guillochon}, E.~{Berger}, B.~D. {Metzger}, P.~S.
  {Cowperthwaite}, M.~{Nicholl}, K.~D. {Alexander}, P.~K. {Blanchard},
  R.~{Chornock}, T.~{Eftekhari}, W.~{Fong}, R.~{Margutti}, P.~K.~G. {Williams},
\newblock \emph{ApJL} \textbf{2017}, \emph{851} L21.

\bibitem{perego17}
A.~{Perego}, D.~{Radice}, S.~{Bernuzzi},
\newblock \emph{ArXiv e-prints} \textbf{2017}.

\bibitem{waxman18}
E.~{Waxman}, E.~O. {Ofek}, D.~{Kushnir}, A.~{Gal-Yam},
\newblock \emph{MNRAS} \textbf{2018}, \emph{481}, 3 3423.

\bibitem{hotokezaka22}
K.~{Hotokezaka}, M.~{Tanaka}, D.~{Kato}, G.~{Gaigalas},
\newblock \emph{arXiv e-prints} \textbf{2022}, arXiv:2204.00737.

\bibitem{hotokezaka20a}
K.~{Hotokezaka}, E.~{Nakar},
\newblock \emph{\apj} \textbf{2020}, \emph{891}, 2 152.

\bibitem{ji19}
A.~P. {Ji}, M.~R. {Drout}, T.~T. {Hansen},
\newblock \emph{\apj} \textbf{2019}, \emph{882}, 1 40.

\bibitem{arnould07}
M.~{Arnould}, S.~{Goriely}, K.~{Takahashi},
\newblock \emph{Phys. Reports} \textbf{2007}, \emph{450} 97.

\bibitem{abadie10}
J.~{Abadie}, B.~P. {Abbott}, R.~{Abbott}, M.~{Abernathy}, T.~{Accadia},
  F.~{Acernese}, C.~{Adams}, R.~{Adhikari}, P.~{Ajith}, B.~{Allen}, et~al.,
\newblock \emph{Classical and Quantum Gravity} \textbf{2010}, \emph{27}, 17
  173001.

\bibitem{abbott21}
R.~{Abbott at al.},
\newblock \emph{arXiv e-prints} \textbf{2021}, arXiv:2111.03634.

\bibitem{dichiara20}
S.~{Dichiara}, E.~{Troja}, B.~{O'Connor}, F.~E. {Marshall}, P.~{Beniamini},
  J.~K. {Cannizzo}, A.~Y. {Lien}, T.~{Sakamoto},
\newblock \emph{\mnras} \textbf{2020}, \emph{492}, 4 5011.

\bibitem{grunthal21}
K.~{Grunthal}, M.~{Kramer}, G.~{Desvignes},
\newblock \emph{\mnras} \textbf{2021}, \emph{507}, 4 5658.

\bibitem{kopparapu08}
R.~K. {Kopparapu}, C.~{Hanna}, V.~{Kalogera}, R.~{O'Shaughnessy},
  G.~{Gonz{\'a}lez}, P.~R. {Brady}, S.~{Fairhurst},
\newblock \emph{\apj} \textbf{2008}, \emph{675}, 2 1459.

\bibitem{mandel22}
I.~{Mandel}, F.~S. {Broekgaarden},
\newblock \emph{Living Reviews in Relativity} \textbf{2022}, \emph{25}, 1 1.

\bibitem{perego19a}
A.~{Perego}, S.~{Bernuzzi}, D.~{Radice},
\newblock \emph{European Physical Journal A} \textbf{2019}, \emph{55}, 8 124.

\bibitem{rosswog14a}
S.~{Rosswog}, O.~{Korobkin}, A.~{Arcones}, F.-K. {Thielemann}, T.~{Piran},
\newblock \emph{MNRAS} \textbf{2014}, \emph{439} 744.

\bibitem{wu22}
Z.~{Wu}, G.~{Ricigliano}, R.~{Kashyap}, A.~{Perego}, D.~{Radice},
\newblock \emph{\mnras} \textbf{2022}, \emph{512}, 1 328.

\bibitem{darbha21}
S.~{Darbha}, D.~{Kasen}, F.~{Foucart}, D.~J. {Price},
\newblock \emph{\apj} \textbf{2021}, \emph{915}, 1 69.

\bibitem{stewart22}
A.~R. {Stewart}, L.-T. {Lo}, O.~{Korobkin}, I.~{Sagert}, J.~{Loiseau},
  H.~{Lim}, M.~A. {Kaltenborn}, C.~M. {Mauney}, J.~{Maxwell Miller},
\newblock \emph{arXiv e-prints} \textbf{2022}, arXiv:2201.01865.

\bibitem{fujibayashi22}
S.~{Fujibayashi}, K.~{Kiuchi}, S.~{Wanajo}, K.~{Kyutoku}, Y.~{Sekiguchi},
  M.~{Shibata},
\newblock \emph{arXiv e-prints} \textbf{2022}, arXiv:2205.05557.

\bibitem{meszaros06}
P.~{Meszaros},
\newblock \emph{Reports of Progress in Physics} \textbf{2006}, \emph{69} 2259.

\bibitem{nakar07}
E.~{Nakar},
\newblock \emph{Phys. Rep.} \textbf{2007}, \emph{442} 166.

\bibitem{lee07}
W.~H. {Lee}, E.~{Ramirez-Ruiz},
\newblock \emph{New Journal of Physics} \textbf{2007}, \emph{9} 17.

\bibitem{kumar15}
P.~{Kumar}, B.~{Zhang},
\newblock \emph{Phys. Rep.} \textbf{2015}, \emph{561} 1.

\bibitem{moesta20}
P.~{M{\"o}sta}, D.~{Radice}, R.~{Haas}, E.~{Schnetter}, S.~{Bernuzzi},
\newblock \emph{\apjl} \textbf{2020}, \emph{901}, 2 L37.

\bibitem{usov92}
V.~V. {Usov},
\newblock \emph{Nature} \textbf{1992}, \emph{357} 472.

\bibitem{nagakura14}
H.~{Nagakura}, K.~{Hotokezaka}, Y.~{Sekiguchi}, M.~{Shibata}, K.~{Ioka},
\newblock \emph{ApJL} \textbf{2014}, \emph{784}, 2 L28.

\bibitem{murguia16}
A.~{Murguia-Berthier}, E.~{Ramirez-Ruiz}, G.~{Montes}, F.~{De Colle},
  L.~{Rezzolla}, S.~{Rosswog}, K.~{Takami}, A.~{Perego}, W.~H. {Lee},
\newblock \emph{ArXiv e-prints} \textbf{2016}.

\bibitem{lazzati17}
D.~{Lazzati}, A.~{Deich}, B.~J. {Morsony}, J.~C. {Workman},
\newblock \emph{\mnras} \textbf{2017}, \emph{471}, 2 1652.

\bibitem{gottlieb18}
O.~{Gottlieb}, E.~{Nakar}, T.~{Piran},
\newblock \emph{\mnras} \textbf{2018}, \emph{473} 576.

\bibitem{lyman18}
J.~D. {Lyman}, G.~P. {Lamb}, A.~J. {Levan}, I.~{Mandel}, N.~R. {Tanvir},
  S.~{Kobayashi}, B.~{Gompertz}, J.~{Hjorth}, A.~S. {Fruchter}, T.~{Kangas},
  D.~{Steeghs}, I.~A. {Steele}, Z.~{Cano}, C.~{Copperwheat}, P.~A. {Evans},
  J.~P.~U. {Fynbo}, C.~{Gall}, M.~{Im}, L.~{Izzo}, P.~{Jakobsson},
  B.~{Milvang-Jensen}, P.~{O'Brien}, J.~P. {Osborne}, E.~{Palazzi}, D.~A.
  {Perley}, E.~{Pian}, S.~{Rosswog}, A.~{Rowlinson}, S.~{Schulze}, E.~R.
  {Stanway}, P.~{Sutton}, C.~C. {Th{\"o}ne}, A.~{de Ugarte Postigo}, D.~J.
  {Watson}, K.~{Wiersema}, R.~A.~M.~J. {Wijers},
\newblock \emph{Nature Astronomy} \textbf{2018}, \emph{2} 751.

\bibitem{mooley18}
K.~P. {Mooley}, A.~T. {Deller}, O.~{Gottlieb}, E.~{Nakar}, G.~{Hallinan},
  S.~{Bourke}, D.~A. {Frail}, A.~{Horesh}, A.~{Corsi}, K.~{Hotokezaka},
\newblock \emph{ArXiv e-prints} \textbf{2018}.

\bibitem{nativi21a}
L.~{Nativi}, M.~{Bulla}, S.~{Rosswog}, C.~{Lundman}, G.~{Kowal}, D.~{Gizzi},
  G.~P. {Lamb}, A.~{Perego},
\newblock \emph{MNRAS} \textbf{2021}, \emph{500}, 2 1772.

\bibitem{nativi21b}
L.~{Nativi}, G.~P. {Lamb}, S.~{Rosswog}, C.~{Lundman}, G.~{Kowal},
\newblock \emph{arXiv e-prints} \textbf{2021}, arXiv:2109.00814.

\bibitem{murguia21}
A.~{Murguia-Berthier}, E.~{Ramirez-Ruiz}, F.~{De Colle}, A.~{Janiuk},
  S.~{Rosswog}, W.~H. {Lee},
\newblock \emph{\apj} \textbf{2021}, \emph{908}, 2 152.

\bibitem{kasen15a}
D.~{Kasen}, R.~{Fern{\'a}ndez}, B.~D. {Metzger},
\newblock \emph{MNRAS} \textbf{2015}, \emph{450} 1777.

\bibitem{wollaeger18a}
R.~T. {Wollaeger}, O.~{Korobkin}, C.~J. {Fontes}, S.~K. {Rosswog}, W.~P.
  {Even}, C.~L. {Fryer}, J.~{Sollerman}, A.~L. {Hungerford}, D.~R. {van
  Rossum}, A.~B. {Wollaber},
\newblock \emph{MNRAS} \textbf{2018}, \emph{478}, 3 3298.

\bibitem{heinzel21}
J.~{Heinzel}, M.~W. {Coughlin}, T.~{Dietrich}, M.~{Bulla}, S.~{Antier},
  N.~{Christensen}, D.~A. {Coulter}, R.~J. {Foley}, L.~{Issa}, N.~{Khetan},
\newblock \emph{\mnras} \textbf{2021}, \emph{502}, 2 3057.

\bibitem{grossman14a}
D.~{Grossman}, O.~{Korobkin}, S.~{Rosswog}, T.~{Piran},
\newblock \emph{MNRAS} \textbf{2014}, \emph{439} 757.

\bibitem{darbha20}
S.~{Darbha}, D.~{Kasen},
\newblock \emph{\apj} \textbf{2020}, \emph{897}, 2 150.

\bibitem{li21}
X.~{Li}, D.~M. {Siegel},
\newblock \emph{\prl} \textbf{2021}, \emph{126}, 25 251101.

\bibitem{bulla21}
M.~{Bulla}, K.~{Kyutoku}, M.~{Tanaka}, S.~{Covino}, J.~R. {Bruten},
  T.~{Matsumoto}, J.~R. {Maund}, V.~{Testa}, K.~{Wiersema},
\newblock \emph{\mnras} \textbf{2021}, \emph{501}, 2 1891.

\bibitem{zhu18}
Y.~{Zhu}, R.~T. {Wollaeger}, N.~{Vassh}, R.~{Surman}, T.~M. {Sprouse}, M.~R.
  {Mumpower}, P.~{M{\"o}ller}, G.~C. {McLaughlin}, O.~{Korobkin}, T.~{Kawano},
  P.~J. {Jaffke}, E.~M. {Holmbeck}, C.~L. {Fryer}, W.~P. {Even}, A.~J.
  {Couture}, J.~{Barnes},
\newblock \emph{\apjl} \textbf{2018}, \emph{863}, 2 L23.

\bibitem{hotokezaka17a}
K.~{Hotokezaka}, R.~{Sari}, T.~{Piran},
\newblock \emph{MNRAS} \textbf{2017}, \emph{468} 91.

\bibitem{barnes16a}
J.~{Barnes}, D.~{Kasen}, M.-R. {Wu}, G.~{Martinez-Pinedo},
\newblock \emph{ApJ} \textbf{2016}, \emph{829} 110.

\bibitem{hotokezaka20}
K.~{Hotokezaka}, E.~{Nakar},
\newblock \emph{\apj} \textbf{2020}, \emph{891}, 2 152.

\bibitem{waxman19}
E.~{Waxman}, E.~O. {Ofek}, D.~{Kushnir},
\newblock \emph{\apj} \textbf{2019}, \emph{878}, 2 93.

\bibitem{winteler12}
C.~{Winteler},
\newblock Ph.D. thesis, University Basel, CH, \textbf{2012}.

\bibitem{winteler12b}
C.~{Winteler}, R.~{K{\"a}ppeli}, A.~{Perego}, A.~{Arcones}, N.~{Vasset},
  N.~{Nishimura}, M.~{Liebend{\"o}rfer}, F.-K. {Thielemann},
\newblock \emph{ApJL} \textbf{2012}, \emph{750} L22.

\bibitem{thielemann11}
F.-K. {Thielemann}, A.~{Arcones}, R.~{K{\"a}ppeli}, M.~{Liebend{\"o}rfer},
  T.~{Rauscher}, C.~{Winteler}, C.~{Fr{\"o}hlich}, I.~{Dillmann}, T.~{Fischer},
  G.~{Martinez-Pinedo}, K.~{Langanke}, K.~{Farouqi}, K.-L. {Kratz}, I.~{Panov},
  I.~K. {Korneev},
\newblock \emph{Progress in Particle and Nuclear Physics} \textbf{2011},
  \emph{66} 346.

\bibitem{rauscher00}
T.~{Rauscher}, F.-K. {Thielemann},
\newblock \emph{Atomic Data and Nuclear Data Tables} \textbf{2000}, \emph{75}
  1.

\bibitem{moeller95}
P.~M\"oller, J.~R. Nix, W.~D. Myers, W.~J. Swiatecki,
\newblock \emph{At. Data Nucl. Data Tables} \textbf{1995}, \emph{59} 185.

\bibitem{fuller82}
G.~M. {Fuller}, W.~A. {Fowler}, M.~J. {Newman},
\newblock \emph{ApJS} \textbf{1982}, \emph{48} 279.

\bibitem{langanke01}
K.~{Langanke}, G.~{Martinez-Pinedo},
\newblock \emph{Atomic Data and Nuclear Data Tables} \textbf{2001}, \emph{79}
  1.

\bibitem{panov10}
I.~V. {Panov}, I.~Y. {Korneev}, T.~{Rauscher}, G.~{Martinez-Pinedo},
  A.~{Keli{\'c}-Heil}, N.~T. {Zinner}, F.-K. {Thielemann},
\newblock \emph{A \& A} \textbf{2010}, \emph{513} A61.

\bibitem{panov05}
I.~V. {Panov}, E.~{Kolbe}, B.~{Pfeiffer}, T.~{Rauscher}, K.-L. {Kratz}, F.-K.
  {Thielemann},
\newblock \emph{Nuclear Physics A} \textbf{2005}, \emph{747} 633.

\bibitem{freiburghaus99a}
C.~Freiburghaus, J.~Rembges, T.~Rauscher, E.~Kolbe, F.-K. Thielemann, K.-L.
  Kratz, J.~Cowan,
\newblock \emph{ApJ} \textbf{1999}, \emph{516} 381.

\bibitem{fernandez15}
R.~{Fernandez}, E.~{Quataert}, J.~{Schwab}, D.~{Kasen}, S.~{Rosswog},
\newblock \emph{MNRAS} \textbf{2015}, \emph{449} 390.

\end{thebibliography}


\end{document}